\documentclass[lettersize,journal]{IEEEtran}
\usepackage{amsmath,amsfonts,amsthm}
\usepackage{algorithmic}
\usepackage{algorithm}
\usepackage{array}
\usepackage[caption=false,font=normalsize,labelfont=sf,textfont=sf]{subfig}
\usepackage{textcomp}
\usepackage{stfloats}
\usepackage{url}
\usepackage{verbatim}
\usepackage{graphicx}
\usepackage{cite}
\begin{document}

\title{Performance Analysis of Metasurface-based Spatial Multimode Transmission for 6G Wireless Communications}

\author{\IEEEauthorblockN{Ju Yong Lee, \IEEEmembership{Member, IEEE}, Seung-Won Keum, Sang Min Oh, Dang-Oh Kim 
and Dong-Ho Cho, \IEEEmembership{Senior Member, IEEE}}\\
\IEEEauthorblockA{\textit{KAIST Institute for Information Technology Convergence, Korea Advanced Institute of Science and Technology}\\
Daejeon, South Korea \\
jylee1117@kaist.ac.kr, keumsw@kaist.ac.kr, ohsangmin@kaist.ac.kr, dangoh@kaist.ac.kr, dhcho@kaist.ac.kr}

}



\maketitle

\begin{abstract}
In 6th generation wireless communication technology, it is important to utilize space resources efficiently. 
Recently, holographic multiple-input multiple-output (HMIMO) and metasurface technology have attracted attention as technologies that maximize space utilization for 6G mobile communications. 
However, studies on HMIMO communications are still in an initial stage
and its fundamental limits are yet to be unveiled.
It is well known that the Fourier transform relationship can be obtained using a lens in the optical field, 
but research to apply it to the mobile communication field is in the early stages. 
In this paper, we show that the Fourier transform relationship between signals can be obtained 
when two metasurfaces are aligned or unaligned, 
and analyze the transmission and reception power, and the maximum number of spatial multimodes that can be transmitted.
In addition, to reduce transmission complexity, we propose a spatial multimode transmission system using three metasurfaces
and analyze signal characteristics on the metasurfaces.
In numerical results, we provide the performance of spatial multimode transmission
in case of using rectangular and Gaussian signals. 
\end{abstract}



\section{Introduction} \label{section1}
\IEEEPARstart{U}ntil the 3rd generation wireless communication, the wireless capacity increase was mainly achieved through the utilization of frequency resources in the time domain. With the introduction of MIMO technology in 4th generation wireless communications, space resources began to be utilized using less than ten antennas. 
In 5th generation wireless communication, the spatial resources of hundreds of antennas are utilized for beamforming, 
and in 6th generation wireless communication, 
the efficient use of spatial resources is expected to play a key role in increasing capacity\cite{ref1}.
In the 5th generation wireless communication, far-field transmission was mainly considered along with the use of multiple beams, 
where each beam can transmit only two streams using polarized waves. 
However, by using an array antenna sufficiently larger than the wavelength length, many streams can be transmitted even in a Line-of-Sight (LoS) environment in the radiating near-field region or Fresnel region\cite{ref2,ref3}.

5G research focuses on novel transceiver hardware architectures and communication algorithms, particularly extreme massive multiple-input multiple-output (mMIMO) systems. These systems enhance spectral efficiency but require many RF chains, leading to high power consumption and hardware costs\cite{ref4}.
To address these challenges, new technologies like holographic MIMO (HMIMO) systems are emerging. 
HMIMO uses metasurfaces to manipulate electromagnetic waves, offering significant potential for 6G requirements. 
However, the studies on HMIMO communications are still at an initial stage, 
and its fundamental limits remain to be unveiled~\cite{ref5}. 
Metasurfaces can control the phase, magnitude, polarization, frequency, and wavefront shape of electromagnetic waves, playing a crucial role in 6G systems\cite{ref6}.

In this paper, we examine spatial resource utilization in the Fresnel region 
when using a sufficiently large metasurface compared to the length of the wavelength.
It is well known that Fourier transform relationships between signals can be obtained using lenses in the optical field~\cite{ref7}.
We propose various performance analysis results to apply the spatial Fourier transform relationship to the mobile communication environment.
It is shown that the Fourier transform relationship can be obtained through the phase transformation of the metasurface 
when the metasurface is aligned and when it is not aligned.
In addition, through analysis of transmission and reception power, 
we see that if they are aligned, all power can be transmitted, 
but if they are not aligned, power loss may occur depending on the tilt.
Just as the maximum dimension that can be transmitted in the time and frequency domains is given, 
the number of spatial multimodes that can be transmitted is analyzed 
when the size of the metasurface is given in the spatial domain. 
We show that the number of spatial multimodes that can be transmitted may decrease in an unaligned metasurface.

If the transmitted and received signals in space have a Fourier transform relationship, 
digital processing must be performed at the transmitter and receiver 
in order to analyze them through signal processing. 
Therefore, the transmitter and receiver require RF chains, DACs, and ADCs 
at all locations at half-wavelength intervals in space, 
which can greatly increase the complexity of the system.
However, the original signal can be recovered by performing the Fourier transform twice. 
Thus, if three metasurfaces are used, the original signal can be recovered at the receiver
through two Fourier transforms that exist among the metasurfaces.
Therefore, in this paper, we propose a system that can restore the original signal at the receiver through two Fourier transforms 
based on three metasurfaces.
Fourier transform signals are expressed on three metasurfaces, 
and the analysis of total power and maximum number of spatial multimodes that can be transmitted is performed.

Section~\ref{section2} describes the characteristics of signals when two metasurfaces are aligned. 
Section~\ref{section3} explains the characteristics of signals when the two metasurfaces are not aligned. 
In section~\ref{section4}, we propose and analyze a spatial multimode transmission system structure using three aligned metasurfaces. 
In section~\ref{section5}, we analyze a spatial multimode transmission system applicable to general environments 
using three unaligned metasurfaces.
In section~\ref{section6}, we examine the Fourier transform relationship on two metasurfaces and examples of spatial multimode transmission using three metasurfaces through numerical analysis.
Conclusions are made in section~\ref{section7}.


\section{Fourier Transform Relation on Aligned Metasurfaces} \label{section2}

\begin{figure}[!t]
\includegraphics[width=3.4in]{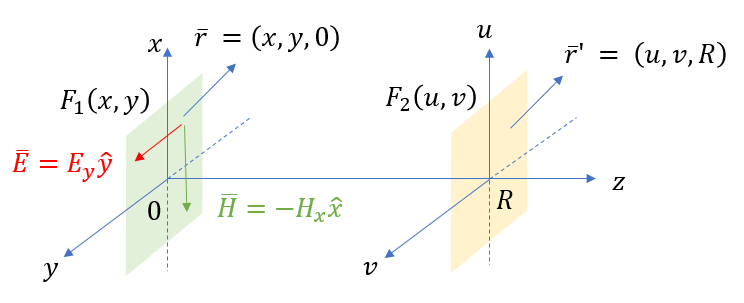}
\caption{Expression of electric (or magnetic) field when transmitting and receiving array antennas are aligned}
\label{fig.1}
\end{figure}

Consider an environment as shown in Fig.~\ref{fig.1}, 
where the distance between the transmitting and receiving antenna arrays is $R$. 
The electric field (or magnetic field) at the transmitting array antenna position, $\bar{r} = ( x,y,0 )$ is expressed as $F_1 (x,y)$, 
and the field at the receiving array antenna position, $\bar{r}' = (u,v,R)$ is expressed as $F_2 (u,v)$.

In this paper, we assume that $\left| x-u \right| \ll R$ and $\left| y-v \right| \ll R$. 
Then, the distance between $\overline{r}$ and $\overline{r}'$ is approximated as follows:
\begin{align}
\left| \overline r - \overline{r}' \right|  & = \sqrt{(x-u)^2+(y-v)^2+R^2 }  \nonumber\\ 
 & = R \left[ 1+\left(\frac{x-u}{R} \right)^2+ \left( \frac{y-v}{R} \right)^2 \right]^{\frac{1}{2}} \nonumber\\
 & \approx\ R \left[ 1+\frac{1}{2} \left( \frac{x-u}{R} \right)^2+\frac{1}{2} \left( \frac{y-v}{R} \right)^2 \right] \nonumber\\ 
 & = R+\frac{ \left( x-u \right)^2}{2R}+\frac{ \left( y-v \right)^2 }{2R}.  \label{eq.1}
\end{align}
Based on Fresnel approximations in \cite{ref7} and \cite{ref8}, 
the relationship between the transmitted and received fields can be expressed as: 
\begin{align}
 F_2\left(u,v\right)
 & = \frac{j}{\lambda}\int_{-\infty}^{\infty} \!\!   \int_{-\infty}^{\infty} \!\!   {F_1(x,y)\ }\frac{e^{-jk\left|\bar{r}-{\bar{r}}^\prime\right|}}{\left|\bar{r}-{\bar{r}}^\prime\right|}dxdy \nonumber\\
 & \approx   \frac{j}{\lambda R} e^{-jkR} \int_{-\infty}^{\infty} \!\!   \int_{-\infty}^{\infty} \!\!   {F_1(x,y)\ } \nonumber\\
 & ~~~~~~~~~~~~~~~~~~~~~~~~~ e^{-\frac{jk}{2R}{(x-u)}^2-\frac{jk}{2R}{(y-v)}^2}dxdy  \nonumber\\
 & = \frac{j}{\lambda R} e^{-jkR} e^{-\frac{jk}{2R}{(u}^2+v^2)} \nonumber\\
 & ~~~\int_{-\infty}^{\infty} \!\!   \int_{-\infty}^{\infty} \!\!   {F_1(x,y)\ }e^{-\frac{jk}{2R}\left(x^2+y^2\right)}e^{\frac{jk}{R}(xu+yv)}dxdy,  \label{eq.2}
\end{align}
where $\lambda$ is the wavelength length, and $k = \frac{2 \pi}{\lambda}$.

Let $S_1 (x,y)$ and $S_2 (u,v)$ be the transmitted signal at $\bar{r} = (x,y,0)$ and the received signal at $\bar{r}' = (u,v,R)$, respectively. 
The transmit and receive fields of $F_1$ and $F_2$ are mapped as follows:
\begin{subequations}
\begin{align} 
S_1(x,y) &= F_1 (x,y) e^{-\frac{jk}{2R}(x^2+y^2)},  \label{eq.3a}\\
S_2(u,v) &= F_2 (u,v) e^{\frac{jk}{2R}(u^2+v^2)}.  \label{eq.3b}
\end{align}
\end{subequations}
Then, we can obtain the following relationship between the transmitted and received signals of $S_1$ and $S_2$:
\begin{equation}
S_2 (u,v) 
\approx \frac{j}{\lambda R} e^{-jkR} 
\int_{-\infty}^{\infty} \!\!   \int_{-\infty}^{\infty} \!\!   
{S_1(x,y)e^{\frac{jk}{R}(xu+yv)}dxdy}.  \label{eq.4}
\end{equation}

\subsection{Fourier transform relationship for signals on aligned metasurfaces}

\begin{figure}[t!]
\centering
\includegraphics[width=2.5in]{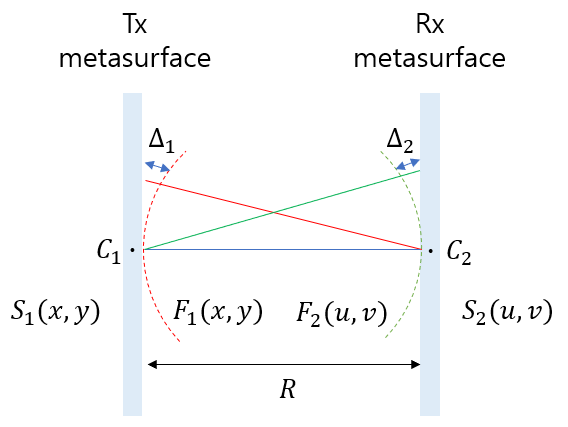}
\caption{Spatial signals on two metasurfaces}
\label{fig.2}
\end{figure}

Consider the environment using two metasurfaces as shown in Fig.~\ref{fig.2}. Metasurfaces exist at the transmitter and the receiver.
Let $S_1$ and $F_1$ be the signal incident on the Tx metasurface and the signal after phase transformation of the Tx metasurface, respectively.
Let $F_2$ and $S_2$ be the signal incident on the Rx metasurface and the signal after phase transformation of the Rx metasurface, respectively.
The transmitter path difference $\Delta_1$ and the receiver path difference $\Delta_2$ in Fig.~\ref{fig.2} can be expressed as follows:
\begin{subequations}
\begin{align}
\Delta_1 
& = \sqrt{ x^2+y^2+R^2 } -R\nonumber\\
&\approx R\left(1+\frac{x^2}{{2R}^2}+\frac{y^2}{{2R}^2}\right)-R = \frac{x^2}{2R}+\frac{y^2}{2R},  \label{eq.5a}  \\ 
\Delta_2 
& = \sqrt{ u^2+v^2+R^2 } - R \approx \frac{{u}^2}{2R}+\frac{v^2}{2R}.   \label{eq.5b}
\end{align}
\end{subequations}
A phase transformation occurs on the Tx metasurface 
so that the plane wave passing through the Tx metasurface is concentrated at the center of the Rx metasurface.
Then, $S_1$ and $F_1$ have the following relationship:
\begin{equation}   
F_1 (x,y) = S_1 (x,y) e^{ j \frac{2\pi}{\lambda} \Delta_1} = S_1 (x,y) e^{ j \frac{k}{2R} \left( x^2 + y^2 \right)}.  \label{eq.6}
\end{equation}
Also, phase conversion is performed on the Rx metasurface 
so that the wave originating from the center of the Tx metasurface can change into a plane wave 
as it passes through the Rx metasurface.
Then, $S_2$ and $F_2$ have the following relationship:
\begin{equation}  
S_2 (u,v) = F_2 (u,v) e^{ j \frac{2\pi}{\lambda} \Delta_2} = F_2 (u,v) e^{ j \frac{k}{2R} \left( u^2 + v^2 \right) }.   \label{eq.7}
\end{equation}
Equations (\ref{eq.6}) and (\ref{eq.7}) are consistent with (\ref{eq.3a}) and (\ref{eq.3b}). 
Therefore, $S_1$ and $S_2$ have the same transform relationship as (\ref{eq.4}) in the transmission and reception environment of Fig.~\ref{fig.2}.

\subsection{Transmit and receive power}
Let's consider the transmit and receive power.
The power $P(S_1)$ of the transmitted signal and the power $P(S_2)$ of the received signal are defined as follows:
\begin{equation}
P(S_i) = \int_{-\infty}^{\infty}  \!\! \int_{-\infty}^{\infty}  \!\! | S_i (x,y) |^2 dx dy,~~i=1,2 .  \label{eq.8}
\end{equation}
From (\ref{eq.4}), 
\begin{align}
P(S_2) 
& = \int_{-\infty}^{\infty}  \!\! \int_{-\infty}^{\infty} \!\! | S_2 (u,v) |^2 du dv \nonumber \\
& \approx \frac{1}{ (\lambda R )^2 } 
\int_{-\infty}^{\infty}  \!\! \int_{-\infty}^{\infty} \!\! \int_{-\infty}^{\infty}  \!\! \int_{-\infty}^{\infty} \!\! S_1 (x,y) S_1^* (x',y')  \nonumber \\
&~~\left( \int_{-\infty}^{\infty}  \!\! \int_{-\infty}^{\infty} \!\! e^{ j \frac{k}{R} (x-x') u + j \frac{k}{R} (y-y') v} du dv \right)
d x' d y' d x d y \nonumber \\
& = \frac{1}{ (\lambda R )^2 } 
\int_{-\infty}^{\infty} \!\! \int_{-\infty}^{\infty} \!\! \int_{-\infty}^{\infty} \!\! \int_{-\infty}^{\infty} \!\! S_1 (x,y) S_1^* (x',y') \nonumber \\
& ~~~ (2 \pi)^2 \delta \left( \frac{k}{R} (x-x') \right) \delta \left( \frac{k}{R} (y-y') \right) dx' dy' d x d y \nonumber \\
& = \int_{-\infty}^{\infty}  \!\! \int_{-\infty}^{\infty}  \!\! | S_1 (x,y) |^2 dx dy  = P(S_1). \label{eq.9}
\end{align}
Therefore, when the Rx metasurface is sufficiently large,
the receive power $P(S_2)$ is approximately equal to the transmit power $P(S_1)$.

\subsection{Maximum number of spatial multimodes}
We introduce the scaled variables and their corresponding functions as follows:
\begin{subequations}
\begin{align} 
& S'_1 (x’,y’) = \frac{1}{ \sqrt{\lambda R} } S_1 (x,y), \label{eq.10a} \\
& S'_2 (u’,v’) = j e^{-jk R} \frac{1}{ \sqrt{\lambda R} } S_2 (u,v),  \label{eq.10b}   
\end{align}
\end{subequations}
where
\begin{subequations}
\begin{align}
&x' = \sqrt{\lambda R} x, ~~ y' = \sqrt{\lambda R} y, \label{eq.11a} \\
&u' = \sqrt{\lambda R} u, ~~ v' = \sqrt{\lambda R} v. \label{eq.11b}           
\end{align}
\end{subequations}
Then,
\begin{subequations}
\begin{align}
&\int_{-\infty}^{\infty} \!\! \int_{-\infty}^{\infty} \!\! | S_1 (x,y) |^2 dx dy = \int_{-\infty}^{\infty} \!\! \int_{-\infty}^{\infty} \!\! | S'_1 (x',y') |^2 dx' dy',  \label{eq.12a} \\
&\int_{-\infty}^{\infty} \!\! \int_{-\infty}^{\infty} \!\! | S_2 (u,v) |^2 du dv = \int_{-\infty}^{\infty} \!\! \int_{-\infty}^{\infty} \!\! | S'_2 (u',v') |^2 du' dv'.  \label{eq.12b}
\end{align}
\end{subequations}
Therefore, $S_1$ and $S'_1$ have the same total power, and $S_2$ and $S'_2$ have the same total power.

By substituting (\ref{eq.10a}), (\ref{eq.10b}), (\ref{eq.11a}), (\ref{eq.11b}) into (\ref{eq.4}),
we can obtain the following Fourier transform relationship between the two signals $S'_1$ and $S'_2$:
\begin{equation} 
S'_2 (u',v')
\approx \int_{-\infty}^{\infty} \!\!   \int_{-\infty}^{\infty} \!\!   
{S'_1 (x',y') e^{ j 2 \pi (x' u' + y' v')}dx' dy'}.  \label{eq.13}
\end{equation}
Therefore, from now on, the relationship in (4) is expressed as having a Fourier transform relationship.

Note that time resource and frequency resource have a Fourier transform relationship. 
Thus, when time and frequency resources are given as $T$ and $W$, respectively, the transmissible dimension becomes $TW$.
Therefore, when the horizontal and vertical sizes of Tx metasurfaces and Rx metasurfaces are $A_{TX}$, $B_{TX}$, $A_{RX}$ and $B_{RX}$, respectively, 
the number of streams $N$ that can be transmitted simultaneously and independently can be expressed as:
\begin{equation}
N 
=\left(\frac{A_{TX}}{\sqrt{\lambda R}}\right)\left(\frac{B_{TX}}{\sqrt{\lambda R}}\right)\left(\frac{A_{RX}}{\sqrt{\lambda R}}\right)\left(\frac{B_{RX}}{\sqrt{\lambda R}}\right)  
=\frac{M_{TX} M_{RX}}{(\lambda R)^2},  \label{eq.14}
\end{equation}
where $M_{TX} = A_{TX}B_{TX}$ and $M_{RX} = A_{RX} B_{RX}$.

\section{Fourier Transform Relation on Unaligned Metasurfaces} \label{section3}

\begin{figure}[t!]
\centering
\includegraphics[width=3.5in]{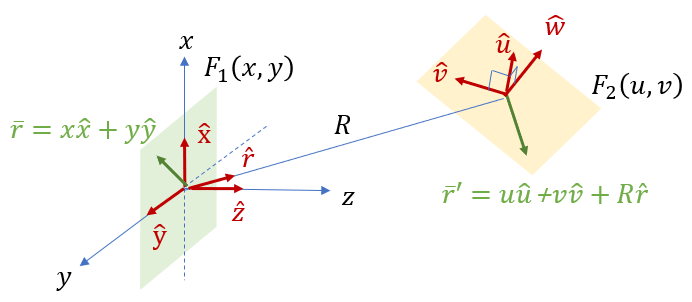}
\caption{Expression of electric (or magnetic) field when transmitting and receiving array antennas are unaligned}
\label{fig.3}
\end{figure}

If the transmit/receive array antennas are not aligned, 
the transmit/receive beamforming that takes into account this condition is required. 
In Fig.~\ref{fig.3}, let the distance between the center of the transmitter and that of the receiver be $R$, 
and the unit direction vector in the direction connecting the centers of the transmitter and receiver is $\hat{r}$. 
Let $\hat{x}$ and $\hat{y}$ be the two-dimensional orthogonal unit vectors constituting the transmitting array antenna plane,
and $\hat{z}$ be the unit vector orthogonal to $\hat{x}$ and $\hat{y}$.
Let $\hat{u}$ and $\hat{v}$ be the two-dimensional orthogonal unit vectors constituting the receiving array antenna plane,
and $\hat{w}$ be the unit vector orthogonal to $\hat{u}$ and $\hat{v}$.
Also, let us express the inner product of several unit vectors as:
\begin{subequations}
\begin{align}
a_{rx} & =\hat{r} \cdot \hat{x},~~ a_{ry} = \hat{r} \cdot \hat{y}, ~~ a_{rz}= \hat{r} \cdot \hat{z}, \label{eq.15a} \\
a_{ru} & =\hat{r} \cdot \hat{u},~~ a_{rv} = \hat{r} \cdot \hat{v}, ~~ a_{rw}= \hat{r} \cdot \hat{w}, \label{eq.15b} \\
a_{xu} & =\hat{x} \cdot \hat{u},~~ a_{yu} =\hat{y} \cdot \hat{u}, \label{eq.15c} \\
a_{xv} & =\hat{x} \cdot \hat{v},~~ a_{yv} = \hat{y} \cdot \hat{v}. \label{eq.15d}
\end{align}
\end{subequations}

Let $F_1 (x,y)$ be the field at the transmitting array antenna position, $\overline{r}=x \hat{x} + y \hat{y}$, 
and $F_2 (u,v)$ be the field at the receiving array antenna position, $\overline{r}' = u \hat{u} + v \hat{v} + R \hat{r}$. 
Then, the square of the distance between $\bar{r}$ and $\bar{r}'$ can be expressed as:
\begin{align}
|\overline{r}-\overline{r}'|^2
& = \left| x \hat{x} +y \hat{y} - u \hat{u} - v \hat{v} -R \hat{r} \right|^2 \nonumber\\
& = R^2+x^2+y^2+u^2+v^2-2 a_{xu} xu -2 a_{xv} xv \nonumber\\
& ~~ -2 a_{yu} yu -2 a_{yv} yv -2 a_{rx} Rx - 2 {a}_{ry} Ry \nonumber\\
& ~~ -2 a_{ru} Ru - 2  a_{rv} Rv  \nonumber \\
& = R^2 + AR + B^2,  \label{eq.16}
\end{align}
where
\begin{subequations}
\begin{align}
A &= -2 a_{rx} x - 2 a_{ry} y + 2 a_{ru} u + 2 a_{rv} v, \label{eq.17a} \\
B^2 &= x^2+y^2+u^2+v^2- 2 a_{xu} xu \nonumber\\
  &~~ -2 a_{xv} xv - 2 a_{yu} yu - 2 a_{yv} yv. \label{eq.17b}
\end{align}
\end{subequations}
Considering (\ref{eq.14}),
$\lambda R$ is the scale of the array antenna area.
Thus, $\frac{A^2}{\lambda R}$ and $\frac{B^2}{\lambda R}$ cannot be ignored.
However, we assume that $|x-u| \ll R$ and $|y-v| \ll R$. Therefore, the followings hold:
\begin{subequations}
\begin{align}
&\frac{A}{R} \ll 1, ~~\frac{B}{R} \ll 1,  \label{eq.18a} \\ 
&\frac{A^{n+1}}{\lambda R^n} \ll 1,~~ \frac{B^{n+1}}{\lambda R^n} \ll 1 ~~ \text{for}~ n \geq 2.  \label{eq.18b}
\end{align}
\end{subequations}
Note that $k |\bar{r} - \bar{r}' |$ is the phase value.
Therefore, the phase term can be approximated as:
\begin{align}
k | \overline{r}-\overline{r}'|
& = \frac{2 \pi R}{\lambda}  \left( 1 + \frac{A}{R} + \frac{B^2}{R^2} \right)^{\frac{1}{2}}  \nonumber\\
& \approx \frac{2 \pi R}{\lambda} \left[ 1 + \frac{1}{2} \left( \frac{A}{R} + \frac{B}{R^2} \right)  - \frac{1}{8} \left( \frac{A}{R} + \frac{B^2}{R^2} \right)^{2}    \right]  \nonumber\\
& \approx \frac{2 \pi R}{\lambda} \left[ 1 + \frac{1}{2} \left( \frac{A}{R} + \frac{B^2}{R^2} \right)  - \frac{1}{8} \frac{A^2}{R^2}  \right]  \nonumber\\
& = \frac{2 \pi R}{\lambda} + \frac{\pi A}{\lambda} + \frac{\pi B^2}{\lambda R} - \frac{\pi A^2}{4 \lambda R}. \label{eq.19}
\end{align}
Substituting (\ref{eq.17a}) and (\ref{eq.17b}) into (\ref{eq.19}), 
\begin{align}
| \overline{r}-\overline{r}'|
& \approx R + \frac{A}{2} + \frac{B^2}{2R} - \frac{A^2}{8R} \nonumber \\
& = R - ( a_{rx} x + a_{ry} y ) + ( a_{ru} u + a_{rv} v ) \nonumber \\
& ~~ + \frac{1}{2R} \left[ x^2 + y^2 - ( a_{rx} x + a_{ry} y )^2  \right]  \nonumber \\
& ~~ + \frac{1}{2R} \left[ u^2 + v^2 - ( a_{ru} u + a_{rv} v )^2   \right]  \nonumber \\
& ~~ - \frac{1}{R} (a_{xu} - a_{rx} a_{ru}) xu - \frac{1}{R} ( a_{xv} - a_{rx} a_{rv} ) xv  \nonumber \\
& ~~ - \frac{1}{R} (a_{yu} - a_{ry} a_{ru} ) yu - \frac{1}{R} ( a_{yv} - a_{ry} a_{rv} ) yv. \label{eq.20}
\end{align}

In the unaligned array antennas, 
modifying (\ref{eq.2}), the relationship between $F_1$ and $F_2$ is as follows:
\begin{equation}
F_2(u,v)
=\frac{j}{\lambda} \int_{-\infty}^{\infty} \!\! \int_{-\infty}^{\infty} \!\!   F_1(x,y)\frac{e^{-jk | \overline{r}-\overline{r}' | } }{ | \overline{r}-\overline{r}' | } a_{rz} a_{rw} dxdy,   \label{eq.21}
\end{equation}
where $a_{rz}$ reflects the degree with which the transmitting array antenna is tilted to the transmission axis, 
and $a_{rw}$ reflects the degree with which the receiving array antenna is tilted to the transmission axis.
Substituting (\ref{eq.20}) into (\ref{eq.21}), we get:
\begin{align}
F_2(u,v)
& \approx\frac{ j }{ \lambda R } e^{-jkR}  
e^{-\frac{jk}{2R}\left[ u^2+v^2 - ( a_{ru} u + a_{rv} v )^2 \right] } \nonumber\\
& ~~ \int_{-\infty}^{\infty} \!\! \int_{-\infty}^{\infty} \!\! 
F_1(x,y) e^{jk( a_{rx} x + a_{ry} y)} \nonumber \\
& ~~~~~e^{-\frac{jk}{2R} \left[ x^2+y^2 - ( a_{rx} x + a_{ry} y )^2 \right] }  \nonumber\\
& ~~~~~ e^{j\frac{k}{R}( b^r_{xu} xu + b^r_{yu} yu + b^r_{xv} xv + b^r_{yv} yv)} a_{rz} a_{rw} dxdy,   \label{eq.22}
\end{align}
where
\begin{subequations}
\begin{align}
&b^r_{xu} = a_{xu} - a_{rx} a_{ru}, ~~ b^r_{xv} = a_{xv} - a_{rx} a_{rv},  \label{eq.23a} \\
&b^r_{yu} = b_{yu} - a_{ry} a_{ru}, ~~ b^r_{yv} = a_{yv} - a_{ry} a_{rv}.  \label{eq.23b}
\end{align}
\end{subequations}

Suppose the transmitting and receiving signals are $S_1$ and $S_2$, respectively. We map two signals of $S_1$ and $S_2$ to $F_1$ and $F_2$ as follows:
\begin{subequations}
\begin{align}  
S_1(x,y)= 
& F_1 (x,y) e^{jk ( a_{rx} x + a_{ry} y)} 
e^{-j\frac{k}{2R} \left[ x^2+y^2 - ( a_{rx} x +  a_{ry} y )^2 \right] }, \label{eq.24a}\\ 
S_2(u,v)= 
& F_2 (u,v) e^{jk ( a_{ru} u + a_{rv} v )} 
e^{j\frac{k}{2R} \left[ u^2+v^2 - ( a_{ru} u + a_{rv} v )^2 \right] }. \label{eq.24b}
\end{align}
\end{subequations}
Then, the two signals $S_1$ and $S_2$ have the following relationship:
\begin{align}
S_2(u,v) 
& \approx \frac{{je}^{jkR}}{\lambda R}  \int_{-\infty}^{\infty} \!\! \int_{-\infty}^{\infty} \!\! {S_1(x,y)}  \nonumber\\  
& e^{j\frac{k}{R}( b^r_{xu} xu + b^r_{yu} yu + b^r_{xv} xv + b^r_{yv} yv )} a_{rz} a_{rw} dxdy.  \label{eq.25}
\end{align}

Now, convert the $(x,y)$ coordinates to the $(x',y')$ coordinates as follows:
\begin{equation}  
\left( \begin{matrix} x' \\ y' \\ \end{matrix} \right) = T \left( \begin{matrix} x\\y \\ \end{matrix} \right),  \label{eq.26}
\end{equation}
where
\begin{equation}
~~ T=\left( \begin{matrix} b^r_{xu} & b^r_{yu} \\ b^r_{xv} & b^r_{yv} \\ \end{matrix} \right). \label{eq.27}
\end{equation}
Also, we define $S'_1 (x', y')$ as follows:
\begin{align}
S_1' (x',y') 
& = \sqrt{  \left| \frac{\partial (x, y)}{ \partial (x',y')}   \right|   }  S_1 (x,y) \nonumber \\
& = \frac{1}{ \sqrt{ || T || } } S_1 (x,y) = \frac{1}{ \sqrt{ || T || } } S_1 ( T^{-1} (x',y') ),   \label{eq.28}
\end{align}
where 
\begin{equation}
|| T || = | b^r_{xu} b^r_{yv} - b^r_{yu} b^r_{xv} |.  \label{eq.29}
\end{equation}
Then, the following holds:
\begin{equation}  
| S_1 (x,y) |^2  dx dy = | S'_1 (x',y') |^2 d x' d y'.  \label{eq.30}
\end{equation}
Applying (\ref{eq.26}) and (\ref{eq.28}) to (\ref{eq.25}), 
we get the Fourier transform relationship between $S'_1$ and $S_2$ as follows:
\begin{align}  
S_2(u,v)
&\approx \frac{{je}^{jkR}}{\lambda R}\frac{a_{rz} a_{rw}}{\sqrt{|| T ||}}    \nonumber \\     
&\int_{-\infty}^{\infty} \!\! \int_{-\infty}^{\infty} \!\! { S'_1 (x',y') e^{j \frac{k}{R} (u x' + v y')}d x' d y'}.  \label{eq.31}
\end{align}
Therefore, the relationship between $S'_1$ and $S_2$ in the unaligned state can be obtained 
in the form similar to the relationship between $S_1$ and $S_2$ in (\ref{eq.4}) in the aligned state.

\subsection{Fourier transform relationship for signals in case of unaligned metasurfaces}

\begin{figure}[t!]
\centering
\includegraphics[width=3in]{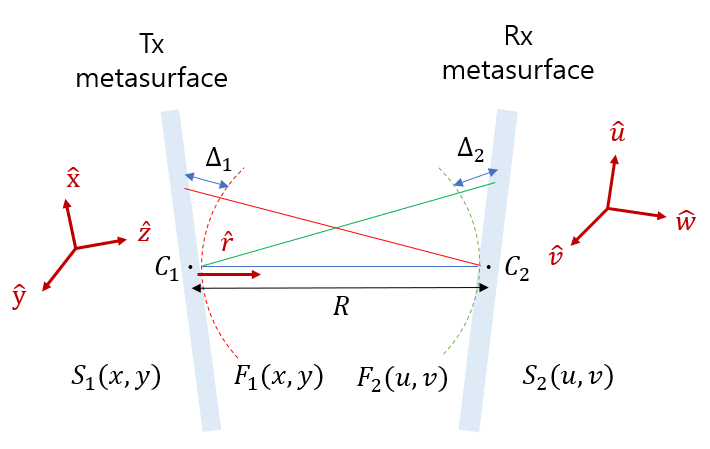}
\caption{Spatial signals on two unaligned metasurfaces}
\label{fig.4}
\end{figure}

Consider an environment using two unaligned metasurfaces at the transmitter and the receiver as shown in Fig.~\ref{fig.4}. 
Let $S_1$ and $F_1$ be the signal incident on the Tx metasurface and the signal after phase transformation of the Tx metasurface, respectively.
Let $F_2$ and $S_2$ be the signal incident on the Rx metasurface and the signal after phase transformation of the Rx metasurface, respectively.
The transmitter path difference $\Delta_1$ and the receiver path difference $\Delta_2$ in Fig.~\ref{fig.4} can be expressed as follows:
\begin{subequations}
\begin{align}
\Delta_1 
& = | x \hat{x} + y \hat{y} - R \hat{r} | -R  \nonumber\\
& = \left[ x^2 + y^2 + R^2 - 2 a_{rx} Rx - 2 a_{ry} Ry  \right]^{\frac{1}{2}} - R   \nonumber \\
& = R \left[ 1 - a_{rx} \frac{2x}{R}  - a_{ry}  \frac{2y}{R} + \frac{x^2 + y^2}{R^2}   \right]^{\frac{1}{2}} - R \nonumber \\
& \approx - a_{rx} x - a_{ry} y + \frac{1}{2R} \left[ x^2 + y^2 - ( a_{rx} x + a_{ry} y )^2 \right], \label{eq.32a} \\
\Delta_2 
& = | u \hat{u} + v \hat{v} - R \hat{r} | - R \nonumber \\
& \approx  a_{ru} u + a_{rv} v + \frac{1}{2R} \left[ u^2 + v^2 - ( a_{ru} u + a_{rv} v )^2 \right]. \label{eq.32b}
\end{align}
\end{subequations}
A phase transformation is performed on the Tx metasurface 
so that plane waves incident perpendicularly on the Tx metasurface converge to the center $C_2$ of the Rx metasurface.
Then, $S_1$ and $F_1$ have the following relationship:
\begin{align}   
F_1 (x,y) & = S_1 (x,y) e^{ j \frac{2\pi}{\lambda} \Delta_1} \nonumber \\
& \approx S_1 (x,y) 
e^{ - jk ( a_{rx} x + a_{ry} y ) }
e^{ j \frac{k}{2R} \left[ x^2 + y^2 - ( a_{rx} x + a_{ry} y)^2 \right]  }. \label{eq.33}
\end{align}
Also, the phase transformation of the Rx metasurface is performed 
so that the wave radiated from the center $C_1$ of the Tx metasurface can be converted into a plane wave perpendicular to the Rx metasurface.
Then, $S_2$ and $F_2$ have the following relationship:
\begin{align}  
S_2 (u,v) & = F_2 (u,v) e^{ j \frac{2\pi}{\lambda} \Delta_2} \nonumber \\
& \approx F_2 (u,v) 
e^{ jk ( a_{ru} u + a_{rv} v ) } 
e^{ j \frac{k}{2R} \left[ u^2 + v^2 - ( a_{ru} u + a_{rv} v)^2 \right] }.  \label{eq.34}
\end{align}
Equations (\ref{eq.33}) and (\ref{eq.34}) are consistent with (\ref{eq.24a}) and (\ref{eq.24b}). 
Therefore, $S_1$ and $S_2$ have the same transform relationship as (\ref{eq.25}) in the transmission and reception environment of Fig.~\ref{fig.4}.

\subsection{Transmit and receive power}

Using (\ref{eq.30}), we can show that the source signal power $P(S'_1)$ is equal to the Tx metasurface signal power $P(S_1)$ as follows:
\begin{align}
P(S'_1) 
&= \int_{-\infty}^{\infty} \!\!    \int_{-\infty}^{\infty} \!\!    | S'_1 (x',y') |^2 dx' dy' \nonumber \\
& = \int_{-\infty}^{\infty} \!\!   \int_{-\infty}^{\infty} \!\!    | S_1 (x,y) |^2 dx dy
= P(S_1).    \label{eq.35}
\end{align}

From (\ref{eq.9}) and (\ref{eq.31}), 
we can obtain the relationship between the source signal power $P(S'_1)$ and the Rx metasurface power $P(S_2)$ as follows:
\begin{align}
P(S_2) 
& = \int_{-\infty}^{\infty} \!\!   \int_{-\infty}^{\infty} \!\!    | S_2 (u,v) |^2 du dv \nonumber \\
& \approx \frac{| a_{rz} a_{rw} |^2 }{ || T || } 
\int_{-\infty}^{\infty} \!\!   \int_{-\infty}^{\infty} \!\!    | S'_1 (x',y') |^2 dx' dy; \nonumber \\
& = \frac{| a_{rz} a_{rw} |^2 }{ || T || } P(S'_1). \label{eq.36}
\end{align}
Also, we can show the following relationship:
\begin{equation}
|| T || = | a_{rz} a_{rw} |.     \label{eq.37}
\end{equation}
The proof of (\ref{eq.37}) is provided in Appendix.
Thus, the following holds:
\begin{align}
P ( S_2 ) \approx || T || P ( S'_1 ) \approx || T || P(S_1).   \label{eq.38}
\end{align}
Suppose either Tx metasurface or Rx metasurface is parallel to the axis direction $\hat{r}$. 
Then, $a_{rz} = 0$ or $a_{rw} = 0$. Therefore, from (\ref{eq.37}) and (\ref{eq.38}), $|| T || = 0$ and the received power $P(S_2)$ is 0. 
On the other hand, if both Tx and Rx metasurfaces are orthogonal to the axis direction $\hat{r}$, 
then $a_{rz} = 1$ and $a_{rw} = 1$, which means that $|| T || = 1$, and we can get the same result as derived on the aligned metasurfaces.

\subsection{Maximum number of spatial multimodes}        
Given a matrix $T$ representing the misalignment state of the Tx and Rx metasurfaces, the number of spatial multiplexing modes, $N$ that can be simultaneously and independently transmitted is expressed as follows:
\begin{equation} 
 N = \frac{M_{TX} M_{RX}}{(\lambda R)^2} || T ||,  \label{eq.39}
\end{equation}
where $M_{TX}$ and $M_{RX}$ are the areas of the Tx and Rx metasurfaces, respectively.
Therefore, the number of spatial multimodes that can be transmitted 
can depend on the angle between the Tx metasurface and the transmission axis, 
and the angle between the Rx metasurface and the transmission axis.


\section{Spatial Multimode Systems using Aligned Metasurfaces} \label{section4}

To transmit spatial multimode using two metasurfaces, 
complex digital processing is required to analyze the Fourier transformed signal spatially. 
However, applying the Fourier transform twice makes the received signal identical to the transmitted signal.
Thus, if three metasurfaces are used, a spatial multimode system can be implemented with low complexity. 
Therefore, in this and the next sections, 
we propose a low-complexity spatial multimode transmission system 
by applying the Fourier transform to the signal twice using three metasurfaces.
This section covers the design for aligned metasurfaces.

\subsection{Spatial multimode transmission using three metasurfaces}

\begin{figure}[t!]
\centering
\includegraphics[width=3.5in]{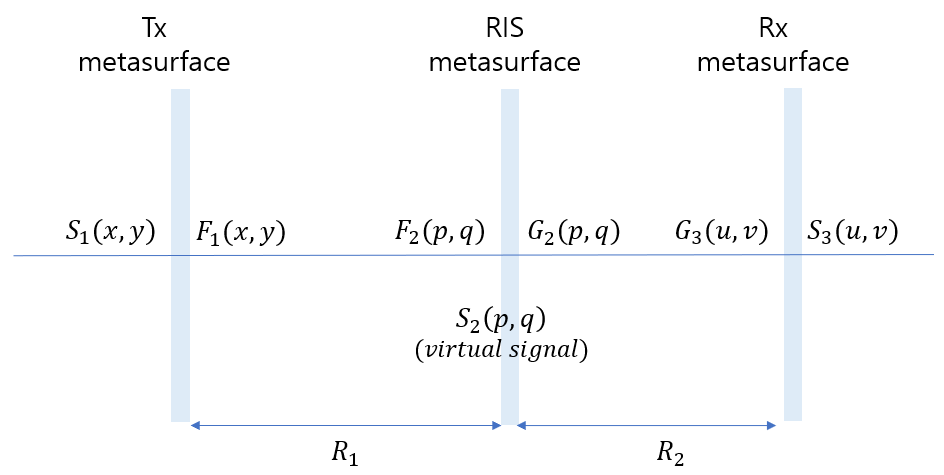}
\caption{Spatial multimode transmission using three aligned metasurfaces}
\label{fig.5}
\end{figure}

Let us consider the environment of three metasurfaces as shown in Fig.~\ref{fig.5}. 
Let the distance between Tx metasurface and RIS metasurface be $R_1$, and the distance between RIS metasurface and Rx metasurface be $R_2$.
The input and output signals of Tx metasurface are $S_1$ and $F_1$, respectively, 
and the input and output signals of RIS metasurface are $F_2$ and $G_2$, respectively. 
Let the input and output signals of Rx metasurface be $G_3$ and $S_3$, respectively.

Considering the relationship between the fields in (\ref{eq.2}), the relationship between $F_1$ and $F_2$ is:
\begin{align}
F_2\left(p,q\right)
&\approx \frac{je^{-jkR_1}}{\lambda R_1}e^{-\frac{jk}{2R_1}{(p^2+q^2)}} \nonumber\\
&\int_{-\infty}^{\infty} \!\!   \int_{-\infty}^{\infty} \!\!   {F_1(x,y)\ }e^{-\frac{jk}{2R_1}\left(x^2+y^2\right)}e^{\frac{jk}{R_1}(x p + y q)}dxdy.  \label{eq.40}
\end{align}
Suppose that the phase change in the Tx metasurface is given by:
\begin{equation}  
S_1\left(x,y\right) = F_1\left(x,y\right) e^{-\frac{jk}{2R_1}\left(x^2+y^2\right)}.   \label{eq.41}
\end{equation}
We also assume that the following is satisfied for the virtual signal $S_2$ in the RIS metasurface:
\begin{equation}   
S_2 \left(p,q\right)= F_2\left(p,q\right) e^{\frac{jk}{2R_1}\left(p^2+q^2\right)}.  \label{eq.42}
\end{equation}
Then, from (\ref{eq.40}), (\ref{eq.41}) and (\ref{eq.42}), 
there is the Fourier transform relationship between $S_1$ and $S_2$, which can be seen as:
\begin{equation}  
S_2 \left( p, q \right) \approx \frac{je^{-j{kR}_1}}{\lambda R_1}\int_{-\infty}^{\infty} \!\!   \int_{-\infty}^{\infty} \!\!   {S_1(x,y)\ }e^{\frac{jk}{R_1}(xp+yq)}dxdy.  \label{eq.43}
\end{equation}

Also, the relationship between $G_2$ and $G_3$ can be obtained as:
\begin{align}
G_3\left(u,v\right)
&\approx \frac{je^{-jkR_2}}{\lambda R_2}e^{-\frac{jk}{2R_2}{(u}^2+v^2)} \nonumber\\
& \int_{-\infty}^{\infty} \!\!   \int_{-\infty}^{\infty} \!\!   { G_2 (p,q) }
e^{-\frac{jk}{2R_2}\left(p^2+q^2\right)}e^{\frac{jk}{R_2}(p u+q v)} d p dq.         \label{eq.44}
\end{align}
Let the virtual signal $S_2$ and the output signal $G_2$ in RIS metasurface be given as:
\begin{equation}   
S_2 \left(p,q\right)= G_2 \left(p,q\right) e^{-\frac{jk}{2R_2}\left(p^2+q^2\right)}.  \label{eq.45}
\end{equation}
Combining (\ref{eq.42}) and (\ref{eq.45}), we can obtain the relation between $G_2$ and $F_2$ as:
\begin{equation} 
G_2 \left(p,q\right)   
=e^{\frac{jk}{2}\left(\frac{1}{R_1}+\frac{1}{R_2}\right)\left(p^2+q^2\right)} F_2 \left(p,q\right). \label{eq.46}
\end{equation}
Let the phase change on the Rx metasurface be given by:
\begin{equation}    
S_3 \left(u,v\right)= G_3 \left(u,v\right) e^{\frac{jk}{2R_2}\left(u^2+v^2\right)}. \label{eq.47}
\end{equation}
Then, from (\ref{eq.44}), (\ref{eq.45}) and (\ref{eq.47}), 
we can see the Fourier transform relationship between $S_2$ and $S_3$ as follows:
\begin{equation}      
S_3 \left(u,v\right) \approx \frac{je^{-j{kR}_2}}{\lambda R_2}
\int_{-\infty}^{\infty} \!\!    \int_{-\infty}^{\infty} \!\!   
{S_2 (p,q) }e^{\frac{jk}{R_2}(p u+q v)}dp dq.  \label{eq.48}
\end{equation}

Let $\theta_1$, $\theta_2$, and $\theta_3$ be the phase shift of the Tx/RIS/Rx metasurfaces, respectively. 
Then, from (\ref{eq.41}), (\ref{eq.46}) and (\ref{eq.47}), 
phase shift values at three metasurfaces are derived as follows:
\begin{subequations}
\begin{align}
\theta_1 (x,y) &= \frac{k}{2 R_1} (x^2 + y^2), \label{eq.49a} \\
\theta_2 (p,q) &= \frac{k}{2} \left( \frac{1}{R_1} + \frac{1}{R_2} \right) \left( p^2 + q^2 \right), \label{eq.49b} \\
\theta_3 (u,v) &= \frac{k}{2 R_2} (u^2 + v^2). \label{eq.49c}
\end{align}
\end{subequations}
If phase control according to the position of the three metasurfaces is performed as shown in (\ref{eq.49a}), (\ref{eq.49b}), (\ref{eq.49c}), 
the Fourier transform relationship is established between signals $S_1$ and $S_2$, and between signals $S_2$ and $S_3$ in Fig.~\ref{fig.5}.

\subsection{Transmit and receive power}

Let $P(S_1)$, $P(S_2)$, $P(S_3)$ be the total power of signals $S_1$, $S_2$, $S_3$, respectively.
Using the method like (\ref{eq.9}), we can obtain the following relation based on (\ref{eq.43}) and (\ref{eq.48}):
\begin{align}
P(S_1) \approx P(S_2) \approx P(S_3). \label{eq.50}
\end{align}
Therefore, if the area of the RIS metasurface is sufficiently large, 
the total power of the RIS metasurface and the Rx metasurface becomes the same as the power of the Tx metasurface.

\subsection{Maximum number of spatial multimodes for example signals} \label{section4.C}

\begin{figure}[t!]
\centering
\includegraphics[width=3.4in]{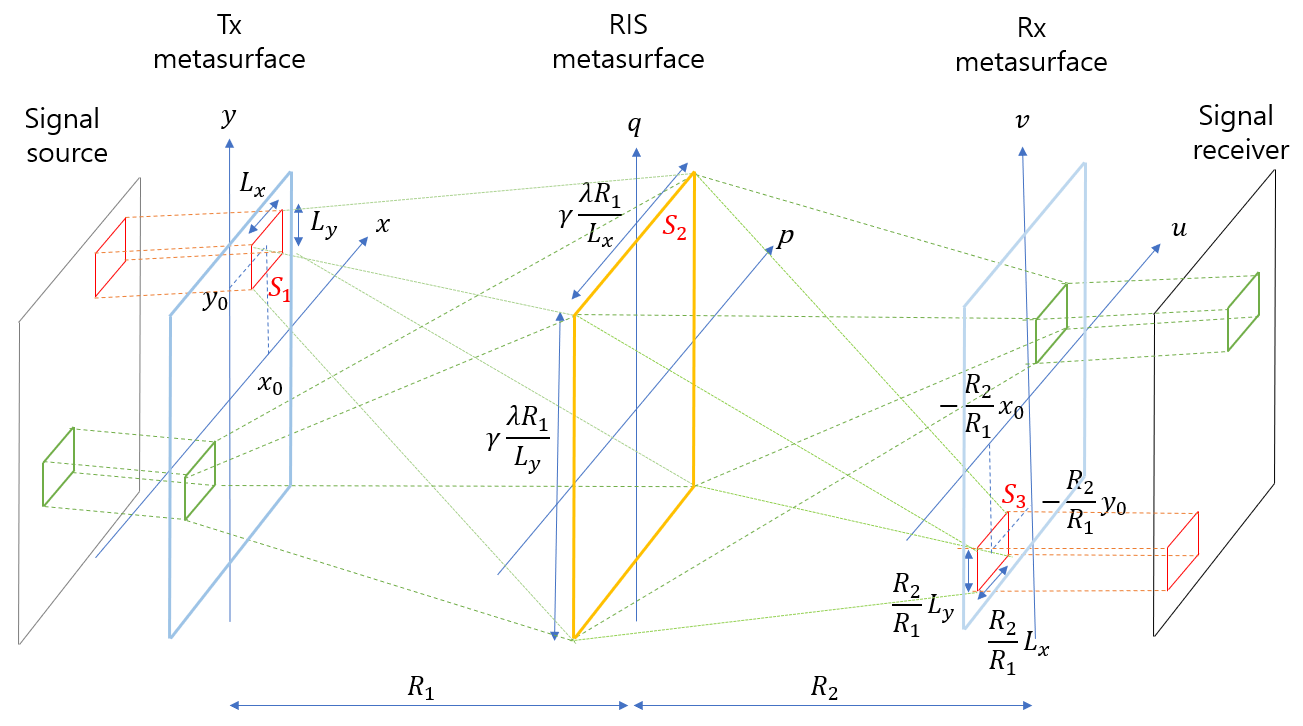}
\caption{Signal areas on the three metasurfaces when a rectangular source signal is transmitted}
\label{fig.6}
\end{figure}

In (\ref{eq.43}) and (\ref{eq.48}), 
$S_1$ and $S_2$ have a Fourier transform relationship, and $S_2$ and $S_3$ have a Fourier transform relationship.
Suppose $S_1$ is given as a rectangular signal as follows:
\begin{equation}
S_1 \left( x, y \right) = \frac{1}{\sqrt{L_x L_y}}  rect\left(\frac{x-x_0}{L_x}\right)rect\left(\frac{y-y_0}{L_y}\right),  \label{eq.51} 
\end{equation}
where
\begin{equation} 
rect(x) = 
   \begin{cases}
       1 ~~~\rm{if}~ - \frac{1}{2} \leq {\it x} \leq \frac{1}{2}, \\ \label{eq.52}
       0 ~~~\rm{otherwise}.       
   \end{cases}
\end{equation}
Then, $S_2$ can be obtained as: 
\begin{align}   
& S_2 \left(p,q \right)
=\frac{ je^{-jkR_1} }{\lambda R_1} \sqrt{L_x L_y} e^{\frac{jk}{R_1}\left(p x_0+q y_0\right)} \nonumber\\
& ~~~~~~~~~~~~~~~~~~~sinc \left(\frac{p L_x}{\lambda R_1}\right) sinc\left(\frac{q L_y}{\lambda R_1}\right),   \label{eq.53}
\end{align}
where $sinc(x) = \frac{sin(\pi x)}{\pi x}$.
Also, $S_3$ can be derived as:
\begin{align}
S_3 \left( u, v \right) & = -\frac{R_1}{R_2} \frac{ e^{-jk\left(R_1+R_2\right)} }{ \sqrt{ L_x L_y } }  \nonumber\\
&rect\left(\frac{u+\frac{R_2}{R_1}x_0}{\frac{R_2}{R_1}L_x}\right) rect\left(\frac{v+\frac{R_2}{R_1}y_0}{\frac{R_2}{R_1}L_y}\right).  \label{eq.54}
\end{align}
Thus, we can obtain the signal power distribution on three metasurfaces as follows:
\begin{subequations}
\begin{align}
& |S_1 (x,y) |^2 =  \frac{1}{L_x L_y} rect \left( \frac{x-x_0}{L_x} \right) rect \left( \frac{y-y_0}{L_y} \right), \label{eq.55a}\\
& |S_2 (p,q) |^2 = \frac{L_x L_y}{ (\lambda R_1)^2 } 
\left[ sinc \left( \frac{p L_x}{\lambda R_1} \right) sinc \left( \frac{q L_y}{\lambda R_1} \right) \right]^2,  \label{eq.55b} \\
& |S_3 (u,v) |^2 = \frac{R_1^2}{L_x L_y R_2^2} rect \left( \frac{u + \frac{R_2}{R_1} x_0 }{ \frac{R_2}{R_1} L_x} \right) rect \left( \frac{v + \frac{R_2}{R_1} y_0}{ \frac{R_2}{R_1} L_y} \right). \label{eq.55c}
\end{align}
\end{subequations}
The signal areas of the Tx metasurface and Rx metasurface are $L_x L_y$ and $L_x L_y \left( \frac{R_2}{R_1} \right)^2$, respectively. 
Suppose that most signal power of RIS metasurface is concentrated in the region where $\left| p \right| < \frac{\gamma}{2} \left( \frac{\lambda R_1}{L_x} \right)$, and $\left| q \right| < \frac{\gamma}{2} \left( \frac{\lambda R_1}{L_y} \right)$.
Fig.~\ref{fig.6} expresses the areas where signals exist on the three metasurfaces.
Let the areas of the three metasurfaces be $M_{TX}$, $M_{RIS}$, and $M_{RX}$, respectively.
Then, the maximum number of spatial multimodes that can be transmitted from Tx to RIS is $\frac{M_{TX}}{L_x L_y}$,
and the maximum number of spatial multimodes that can be transmitted from RIS to Rx is $\frac{M_{RX}}{L_x L_y} \left( \frac{R_1}{R_2} \right)^2$.
Since $M_{RIS}=\frac{(\gamma \lambda R_1)^2}{L_x L_y}$, the maximum number of spatial multimodes, $N$ can be expressed as:
\begin{equation} 
N = \frac{1}{\gamma^2} \min \left( \frac{M_{TX} M_{RIS}}{(\lambda R_1)^2}, \frac{M_{RIS} M_{RX}}{ (\lambda R_2)^2}   \right).  \label{eq.56}
\end{equation}

On the other hand, suppose that $S_1$ is a two-dimensional Gaussian signal with independent variables $x$ and $y$ as follows:
\begin{equation}
S_1 \left( x, y \right) = \frac{1}{\sqrt{ 2 \pi \sigma_x \sigma_y } } 
\exp \left( -  \frac{(x - x_0)^2}{ 4 \sigma_x^2 } - \frac{( y - y_0 )^2}{ 4 \sigma_y^2}  \right).  \label{eq.57} 
\end{equation}
Then, $S_2$ and $S_3$ can be obtained as follows: 
\begin{subequations}
\begin{align}   
S_2 \left(p,q \right)
&= \frac{j e^{-j k R_1}}{\lambda R_1} 2 \sqrt{2 \pi \sigma_x \sigma_y} e^{ \frac{jk}{R_1} ( p x_0 + q y_0 ) } e^{ - \frac{k^2}{R_1^2} ( \sigma_x^2 p^2 + \sigma_y^2 q^2 ) },   \label{eq.58a} \\
S_3 \left( u, v \right) &= 
- \frac{R_1}{R_2} \frac{ e^{-jk (R_1 + R_2)}}{ \sqrt{ 2 \pi \sigma_x \sigma_y } }  \nonumber\\ 
&\exp \left( - \frac{ (u+ \frac{R_2}{R_1} x_0)^2 }{ 4 (\frac{R_2}{R_1} \sigma_x)^2 }    
- \frac{ (v + \frac{R_2}{R_1} y_0 )^2 }{  4 ( \frac{R_2}{R_1} \sigma_y )^2 }  \right).  \label{eq.58b}
\end{align}
\end{subequations}
Thus, we can obtain the signal power distribution on three metasurfaces as follows:
\begin{subequations}
\begin{align}
& |S_1 (x,y) |^2 =  \frac{1}{2 \pi \sigma_x \sigma_y} \exp \left( -  \frac{(x - x_0)^2}{ 2 \sigma_x^2 } - \frac{( y - y_0 )^2}{ 2 \sigma_y^2}  \right) \label{eq.59a}\\
& |S_2 (p,q) |^2 = \frac{8 \pi \sigma_x \sigma_y}{ (\lambda R_1)^2 } 
\exp \left( - \frac{2 k^2}{R_1^2} ( \sigma_x^2 p^2 + \sigma_y^2 q^2 ) \right)  \label{eq.59b} \\
& |S_3 (u,v) |^2 = \frac{1}{2 \pi \sigma_x \sigma_y} \left( \frac{R_1}{R_2} \right)  \nonumber \\
&~~~~~~~~~~~~~\exp \left( - \frac{ (u+ \frac{R_2}{R_1} x_0)^2 }{ 2 (\frac{R_2}{R_1} \sigma_x)^2 }    
- \frac{ (v + \frac{R_2}{R_1} y_0 )^2 }{  2 ( \frac{R_2}{R_1} \sigma_y )^2 }  \right) \label{eq.59c}
\end{align}
\end{subequations}
Note that the powers of all three signals of $S_1$, $S_2$, and $S_3$ have Gaussian distributions.
The mean and standard deviation of the distribution $|S_1|^2$ are $(x_0, y_0)$ and $(\sigma_x,\sigma_y)$, respectively.
The mean and standard deviation of the distribution $|S_2|^2$ is $(0,0)$ and $(\frac{\lambda R_1}{4 \pi \sigma_x},\frac{\lambda R_1}{4 \pi \sigma_y})$, respectively. 
The mean and standard deviation of the distribution $|S_3|^2$ are $\left(-\frac{R_2}{R_1} x_0, -\frac{R_2}{R_1} y_0 \right)$ and $\left( \frac{R_2}{R_1} \sigma_x, \frac{R_2}{R_1} \sigma_y  \right)$, respectively.
Suppose that the signal areas of Tx metasurface and Rx metasurface are $\gamma_1^2 \sigma_x \sigma_y$ and $\gamma_1^2 \sigma_x \sigma_y \left( \frac{R_2}{R_1} \right)^2$, respectively, 
and most signal power of RIS metasurface is concentrated in the region where $\left| p \right| < \frac{\gamma_2}{2} \left( \frac{\lambda R_1}{\sigma_x} \right)$, and $\left| q \right| < \frac{\gamma_2}{2} \left( \frac{\lambda R_1}{\sigma_y} \right)$.
Then, the maximum number of spatial multimodes that can be transmitted from Tx to RIS is $\frac{M_{TX}}{\gamma_1^2 \sigma_x \sigma_y}$,
and the maximum number of spatial multimodes from RIS to Rx is $\frac{M_{RX}}{\gamma_1^2 \sigma_x \sigma_y} \left( \frac{R_1}{R_2} \right)^2$.
Since $M_{RIS}=\frac{(\gamma_2 \lambda R_1)^2}{\sigma_x \sigma_y}$, the maximum number of spatial multimodes, $N$ from Tx to Rx can be expressed as:
\begin{equation}
N = \frac{1}{(\gamma_1 \gamma_2)^2}  \min \left( \frac{M_{TX} M_{RIS}}{(\lambda R_1)^2}, \frac{M_{RIS} M_{RX}}{ (\lambda R_2)^2}   \right).   \label{eq.60}
\end{equation}

Considering (\ref{eq.14}), 
the maximum number of spatial multimodes in (\ref{eq.56}) and (\ref{eq.60}) is reduced 
by $\frac{1}{\gamma^2}$ and $\frac{1}{(\gamma_1 \gamma_2)^2}$ times 
compared to the theoretical maximum number of spatial multimodes, respectively.


\section{Spatial Multimode Systems using Unaligned Metasurfaces} \label{section5}

In this section, when three metasurfaces are unaligned in space, 
two Fourier transform relationships are obtained through the phase change of the metasurfaces, 
and the performance of spatial multimode transmission is analyzed.

\subsection{Spatial multimode transmission using three metasurfaces}

\begin{figure}[t!]
\centering
\includegraphics[width=3.5in]{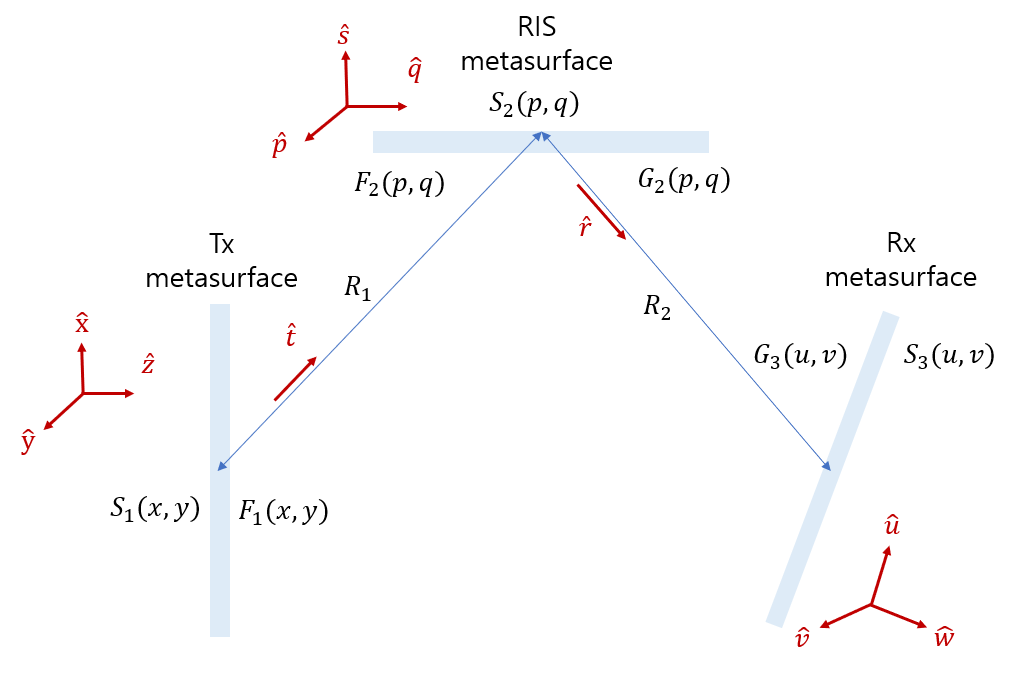}
\caption{Spatial multimode transmission using three unaligned metasurfaces}
\label{fig.7}
\end{figure}

Let us consider the environment of three metasurfaces as shown in Fig.~\ref{fig.7}. 
The Tx metasurface is formed by perpendicular unit vectors $\hat{x}$ and $\hat{y}$, and $\hat{z}$ is a unit vector perpendicular to $\hat{x}$ and $\hat{y}$. 
The RIS metasurface is formed by perpendicular unit vectors $\hat{p}$, $\hat{q}$, and $\hat{s}$ is a unit vector perpendicular to $\hat{p}$ and $\hat{q}$. 
The Rx metasurface is formed by perpendicular unit vectors $\hat{u}$ and $\hat{v}$, and $\hat{w}$ is a unit vector perpendicular to $\hat{u}$ and $\hat{v}$.
Additionally, let $\hat{t}$ be the unit direction vector connecting the center of the Tx metasurface and that of the RIS metasurface, 
and $\hat{r}$ be the unit direction vector connecting the center of the RIS metasurface and that of the Rx metasurface.
Also, let the distance between Tx metasurface and RIS metasurface be $R_1$, and the distance between RIS metasurface and Rx metasurface be $R_2$.
The input and output signals of Tx metasurface are $S_1$ and $F_1$, respectively, 
and the input and output signals of RIS metasurface are $F_2$ and $G_2$, respectively. 
Also, the input and output signals of Rx metasurface are $G_3$ and $S_3$, respectively.

Considering the relationship between the fields in (\ref{eq.22}), the relationship between $F_1$ and $F_2$ is expressed as:
\begin{align}
F_2 (p,q)
& \approx \frac{j e^{-jk R_1}}{\lambda R_1} e^{-\frac{jk}{2R_1} \left[ p^2+q^2 - ( a_{t p} p + a_{t q} q )^2 \right] } \nonumber\\
& \int_{-\infty}^{\infty} \!\!    \int_{-\infty}^{\infty} \!\!    F_1(x,y) e^{jk ( a_{tx} x + a_{ty} y ) }  \nonumber \\
& e^{-\frac{jk}{2R_1} \left[ x^2+y^2 - ( a_{tx} x + a_{ty} y )^2 \right] }  \nonumber \\
& e^{\frac{jk}{R_1} ( b^t_{xp} xp + b^t_{xq} xq + b^t_{yp} yp + b^t_{yq} yq )} a_{tz} a_{t s} dxdy,  \label{eq.61}
\end{align}
where
\begin{subequations}
\begin{align}
&a_{tx}  =\hat{t} \cdot \hat{x},~~ a_{ty} = \hat{t} \cdot \hat{y}, ~~ a_{tz}= \hat{t} \cdot \hat{z}, \label{eq.62a}\\
&a_{tp}  =\hat{t} \cdot \hat{p},~~ a_{tq} = \hat{t} \cdot \hat{q}, ~~ a_{ts}= \hat{t} \cdot \hat{s}, \label{eq.62b}\\
&a_{xp}  =\hat{x} \cdot \hat{p},~~ a_{yp} =\hat{y} \cdot \hat{p}, \label{eq.62c}\\
&a_{xq}  =\hat{x} \cdot \hat{q},~~ a_{yq} = \hat{y} \cdot \hat{q}, \label{eq.62d}
\end{align}
\end{subequations}
and
\begin{subequations}
\begin{align}
&b^t_{xp} = a_{xp} - a_{tx} a_{tp}, ~~ b^t_{xq} = a_{xq} - a_{tx} a_{tq},  \label{eq.63a} \\
&b^t_{yp} = a_{yp} - a_{ty} a_{tp}, ~~ b^t_{yq} = a_{yq} - a_{ty} a_{tq}.  \label{eq.63b}
\end{align}
\end{subequations}
Suppose that the phase change in the Tx metasurface is given by:
\begin{equation}  
S_1 (x,y) = F_1 (x,y) e^{ jk ( a_{tx} x + a_{ty} y )  }
e^{ -\frac{jk}{2R_1} \left[ x^2+y^2 - ( a_{tx} x + a_{ty} y )^2 \right]}.    \label{eq.64}
\end{equation}
We also assume that the following is satisfied for the virtual signal $S_2$ in the RIS metasurface:
\begin{equation}   
S_2 (p,q) = F_2 (p,q) 
e^{ jk ( a_{tp} p + a_{tq} q )}  
e^{ \frac{jk}{2R_1} \left[ p^2 + q^2 
- ( a_{tp} p + a_{tq} q )^2 \right]}.   \label{eq.65}
\end{equation}
Then, from (\ref{eq.61}), (\ref{eq.64}) and (\ref{eq.65}), 
we can derive the relationship between $S_1$ and $S_2$, which can be seen as:
\begin{align}   
S_2 \left( p, q \right) 
& \approx \frac{je^{-j{kR}_1}}{\lambda R_1} 
\int_{-\infty}^{\infty} \!\!    \int_{-\infty}^{\infty} \!\!    S_1(x,y)  \nonumber \\
&e^{\frac{jk}{R_1} ( b^t_{xp} xp + b^t_{xq} xq + b^t_{yp} yp + b^t_{yq} yq )} a_{tz} a_{t s} dx dy.
\label{eq.66}
\end{align}

Also, the relationship between $G_2$ and $G_3$ can be obtained as:
\begin{align}
G_3 (u,v)
&\approx \frac{je^{-jkR_2}}{\lambda R_2} e^{-\frac{jk}{2R_2} \left[ u^2+v^2 - ( a_{ru} u + a_{rv} v )^2 \right] } \nonumber\\
& \int_{-\infty}^{\infty} \!\!    \int_{-\infty}^{\infty} \!\!    G_2 (p,q) e^{ jk ( a_{rp} p + a_{rq} q ) } \nonumber \\
& e^{-\frac{jk}{2R_2}\left[ p^2 + q^2 - ( a_{rp} p + a_{rq} q )^2 \right]}  \nonumber \\
& e^{\frac{jk}{R_2}( b^r_{pu} pu + b^r_{pv} pv + b^r_{qu} qu + b^r_{qv} qv )} a_{rs} a_{rw} dp dq.         \label{eq.67}
\end{align}
where
\begin{subequations}
\begin{align}
&a_{rp}  =\hat{r} \cdot \hat{p},~~ a_{rq} = \hat{r} \cdot \hat{q}, ~~ a_{rs}= \hat{r} \cdot \hat{s}, \label{eq.68a}\\
&a_{ru}  =\hat{r} \cdot \hat{u},~~ a_{rv} = \hat{r} \cdot \hat{v}, ~~ a_{rw}= \hat{r} \cdot \hat{w}, \label{eq.68b}\\
&a_{pu}  =\hat{p} \cdot \hat{u},~~ a_{qu} =\hat{q} \cdot \hat{u}, \label{eq.68c}\\
&a_{pv}  =\hat{p} \cdot \hat{v},~~ a_{qv} = \hat{q} \cdot \hat{v}, \label{eq.68d}
\end{align}
\end{subequations}
and
\begin{subequations}
\begin{align}
&b^r_{pu} = a_{pu} - a_{rp} a_{ru}, ~~ b^r_{pv} = a_{pv} - a_{rp} a_{rv},  \label{eq.69a} \\
&b^r_{qu} = a_{qu} - a_{rq} a_{ru}, ~~ b^r_{qv} = a_{qv} - a_{rq} a_{rv}.  \label{eq.69b}
\end{align}
\end{subequations}
Let the virtual signal $S_2$ have the following relationship with the output signal $G_2$ in the RIS metasurface:
\begin{equation}    
S_2 (p,q)= G_2 (p,q) e^{jk ( a_{rp} p + a_{rq} q ) } 
e^{-\frac{jk}{2R_2} \left[ p^2 + q^2 - ( a_{rp} p + a_{rq} q )^2 \right]}.  \label{eq.70}
\end{equation}
Combining (\ref{eq.65}) and (\ref{eq.70}), we can obtain the relation between $G_2$ and $F_2$ as:
\begin{align} 
G_2 \left(p,q\right)    
& = e^{ jk ( a_{tp} p + a_{tq} q ) } 
    e^{ - jk ( a_{rp} p + a_{rq} q )}  \nonumber \\
& e^{ \frac{jk}{2R_1} \left[ p^2 + q^2 - ( a_{tp} p + a_{tq} q )^2 \right]} \nonumber \\
& e^{\frac{jk}{2R_2} \left[ p^2 + q^2 - ( a_{rp} p + a_{rq} q )^2 \right]} F_2 (p,q). \label{eq.71}
\end{align}
Let the phase change on the Rx metasurface be given by:
\begin{equation}    
S_3 (u,v) = G_3 (u,v) e^{jk ( a_{ru} u + a_{rv} v )} 
e^{ \frac{jk}{2R_2} \left[u^2+v^2 - ( a_{ru} u + a_{rv} v )^2 \right]}.  \label{eq.72}
\end{equation}
Then, from (\ref{eq.67}), (\ref{eq.70}) and (\ref{eq.72}), 
the relationship between $S_2$ and $S_3$ can be derived as follows:
\begin{align}     
S_3 (u,v) 
&\approx \frac{je^{-j{kR}_2}}{\lambda R_2} 
\int_{-\infty}^{\infty} \!\!    \int_{-\infty}^{\infty} \!\!    S_2 (p,q) \nonumber \\
& e^{ \frac{jk}{R_2} ( b^r_{pu} pu + b^r_{pv} pv + b^r_{qu} qu + b^r_{qv} qv ) } a_{rs} a_{rw} dp dq.   \label{eq.73}
\end{align}

Now, we convert the $(x,y)$ coordinates to the $(x',y')$ coordinates as follows:
\begin{equation}  
\left( \begin{matrix} x' \\ y' \\ \end{matrix} \right) = T_1 \left( \begin{matrix} x\\y \\ \end{matrix} \right), \label{eq.74}
\end{equation}
where
\begin{align}
~~ T_1 =\left( \begin{matrix} b^t_{xp} & b^t_{yp} \\ b^t_{xq} & b^t_{yq} \\ \end{matrix} \right),
~~~ || T_1 || = | b^t_{xp} b^t_{yq} -  b^t_{yp} b^t_{xq } |.   \label{eq.75}
\end{align} 
Also, we define $S'_1 (x', y')$ as follows:
\begin{align}
S'_1 (x',y') 
& = \sqrt{  \left| \frac{\partial (x, y)}{ \partial (x',y')}   \right|   }  S_1 (x,y) \nonumber \\
& = \frac{1}{ \sqrt{ ||T_1|| } } S_1 (x,y) = \frac{1}{ \sqrt{ || T_1 || } } S_1 ( T_1^{-1} (x',y') ),   \label{eq.76}
\end{align}
where
\begin{equation}  
| S_1 (x,y) |^2  dx dy = | S'_1 (x',y') |^2 d x' d y'. \label{eq.77}
\end{equation}
Then, applying (\ref{eq.74}) and (\ref{eq.76}) to (\ref{eq.66}), 
we get the Fourier transform relation between $S'_1$ and $S_2$ as follows:
\begin{align}  
S_2(p,q) 
& \approx \frac{{je}^{jkR}}{\lambda R}\frac{a_{tz} a_{t s}}{\sqrt{||T_1||}} \nonumber \\        
& \int_{-\infty}^{\infty} \!\! \int_{-\infty}^{\infty} \!\! {S'_1 (x',y') e^{j\frac{k}{R}(x' p + y' q)}d x' d y'}.   \label{eq.78}
\end{align}
Moreover, the $(u,v)$ coordinates are converted to $(u',v')$ coordinates as follows:
\begin{equation}  
\left( \begin{matrix} u' \\ v' \\ \end{matrix} \right) = T_2 \left( \begin{matrix} u\\v \\ \end{matrix} \right),  \label{eq.79}
\end{equation}
where
\begin{align}
~~ T_2 =\left( \begin{matrix} b^r_{pu} &  b^r_{pv} \\ b^r_{qu} & b^r_{qv} \\ \end{matrix} \right), 
~~ || T_2 || = | b^r_{pu} b^r_{qv} - b^r_{qu} b^r_{pv} |.  \label{eq.80}
\end{align}
Also, we define $S'_3 (u', v')$ as follows:
\begin{align}
S'_3 (u',v') 
& = \sqrt{  \left| \frac{\partial (u, v)}{ \partial (u',v')}   \right|   }  S_3 (u,v) \nonumber \\
& = \frac{1}{ \sqrt{ ||T_2|| } } S_3 (u,v) = \frac{1}{ \sqrt{ ||T_2|| } } S_3 ( T_2^{-1} (u',v') ),   \label{eq.81}
\end{align}
where
\begin{equation} 
| S_3 (u,v) |^2  du dv = | S'_3 (u',v') |^2 d u' d v'.  \label{eq.82}
\end{equation}
Then, applying (\ref{eq.79}) and (\ref{eq.81}) to (\ref{eq.73}), 
we get the Fourier transform relation between $S_2$ and $S'_3$ as follows:
\begin{align}  
S'_3(u',v') 
& \approx \frac{{je}^{jkR}}{\lambda R}\frac{a_{r s} a_{rw}}{\sqrt{||T_2||}}  \nonumber \\     
& \int_{-\infty}^{\infty} \!\! \int_{-\infty}^{\infty} \!\! {S_2 (p,q) e^{j\frac{k}{R}(p u' + q v')}d p d q}.   \label{eq.83}
\end{align}

Let $\theta_1$, $\theta_2$, and $\theta_3$ be the phase shift of the Tx/RIS/Rx metasurfaces, respectively. 
Then, from (\ref{eq.64}), (\ref{eq.71}) and (\ref{eq.72}), phase shift values at three metasurfaces are derived as follows:
\begin{subequations}
\begin{align}
\theta_1 (x,y) 
& = - k ( a_{tx} x + a_{ty} y ) \nonumber \\
& + \frac{k}{2 R_1} \left[ x^2 + y^2 - ( a_{tx} x + a_{ty} y )^2 \right], \label{eq.84a} \\
\theta_2 (p,q)  
& = k ( a_{tp} p + a_{tq} q ) - k ( a_{r p} p + a_{rq} q ) \nonumber \\
& + \frac{k}{2R_1} \left[ p^2 + q^2 - ( a_{tp} p + a_{tq} q )^2 \right]  \nonumber \\
& + \frac{k}{2R_2} \left[ p^2 + q^2 - ( a_{rp} p + a_{rq} q )^2 \right], \label{eq.84b} \\
\theta_3 (u,v) 
& = k ( a_{ru} u + a_{rv} v ) \nonumber \\ 
& + \frac{k}{2 R_2} \left[ u^2 + v^2 - ( a_{ru} u + a_{rv} v )^2 \right]. \label{eq.84c}
\end{align}
\end{subequations}
If phase control according to the position of the three metasurfaces is performed as shown in (\ref{eq.84a}), (\ref{eq.84b}), and (\ref{eq.84c}), 
the Fourier transform relationship is established between signals $S'_1$ and $S_2$ as well as between signals $S_2$ and $S'_3$ in Fig.~\ref{fig.7}.

\subsection{Transmit and receive power}

Let $P(S'_1)$, $P(S_1)$, $P(S_2)$, $P(S_3)$, $P(S'_3)$ be the power of signals $S'_1$, $S_1$, $S_2$, $S_3$, $S'_3$, respectively.
We can obtain the following equations from (\ref{eq.77}) and (\ref{eq.82}):
\begin{subequations}
\begin{align}
&P(S'_1) = P(S_1),    \label{eq.85a} \\
&P(S'_3) = P(S_3).    \label{eq.85b}
\end{align}
\end{subequations}
Also, considering the relation of (\ref{eq.38}), 
we can obtain the following relation from (\ref{eq.78}) and (\ref{eq.83}):
\begin{subequations}
\begin{align}
&P(S_2) \approx ||T_1|| P(S'_1) = || T_1 || P(S_1),    \label{eq.86a} \\
&P(S_3) = P(S'_3) \approx ||T_2 || P(S_2) \approx ||T_1 T_2 || P(S_1).   \label{eq.86b}
\end{align}
\end{subequations}

\subsection{Maximum number of spatial multimodes for example signals}

Consider example signals in section~\ref{section4.C}.
When the input signals are rectangular as shown in (\ref{eq.51}), 
the maximum number of spatial multimodes $N$ in (\ref{eq.56}) can be generalized as follows:
\begin{align} 
N = \frac{1}{\gamma^2} \min \left( \frac{M_{TX} M_{RIS}}{(\lambda R_1)^2} || T_1 || , 
\frac{M_{RIS} M_{RX}}{ (\lambda R_2)^2}  || T_2 ||  \right). \label{eq.87}
\end{align}
When the input signals are Gaussian as shown in (\ref{eq.57}), 
the maximum number of spatial multimodes $N$ in (\ref{eq.60}) can be generalized as follows:
\begin{align} 
N = \frac{1}{(\gamma_1 \gamma_2)^2}  \min \left( \frac{M_{TX} M_{RIS}}{(\lambda R_1)^2} ||T_1||, 
\frac{M_{RIS} M_{RX}}{ (\lambda R_2)^2}  ||T_2|| \right). \label{eq.88}
\end{align}


\section{Numerical Results and Discussions} \label{section6}
This section presents some numerical results for spatial multimode transmission.
We assume that the carrier frequency is $30~GHz$ and the wavelength $\lambda$ is $0.01~m$.

\subsection{Fourier transform relationship for two metasurfaces}
We consider the environment with two metasurfaces.
The distance $R$ between Tx metasurface and Rx metasurface is $10~m$.

\begin{figure*}[!t]
\centering
\subfloat[]{\includegraphics[width=2.4in]{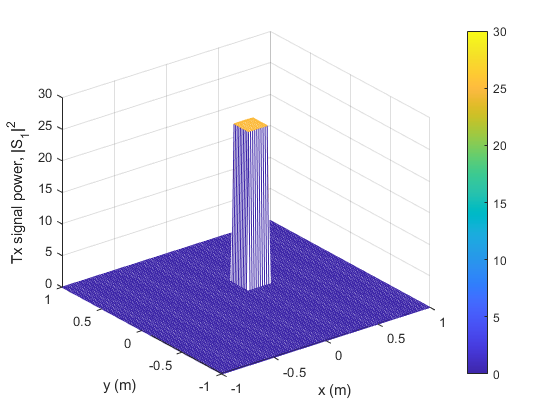}%
\label{fig.8_1}}
\hfil
\subfloat[]{\includegraphics[width=2.4in]{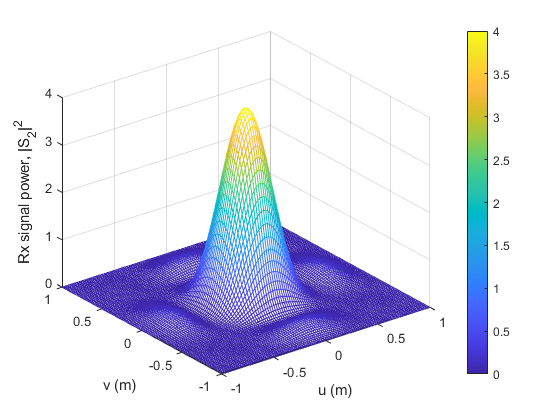}%
\label{fig.8_2}}
\hfil
\subfloat[]{\includegraphics[width=2.4in]{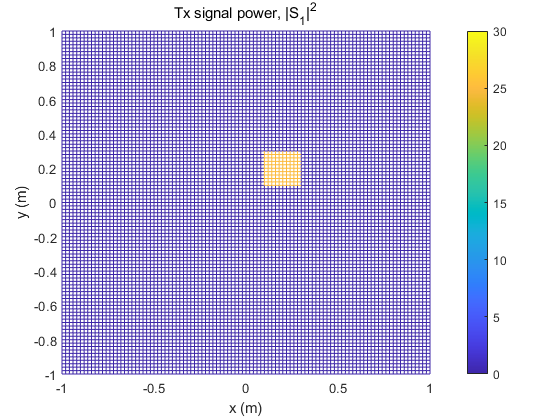}%
\label{fig.8_3}}
\hfil
\subfloat[]{\includegraphics[width=2.4in]{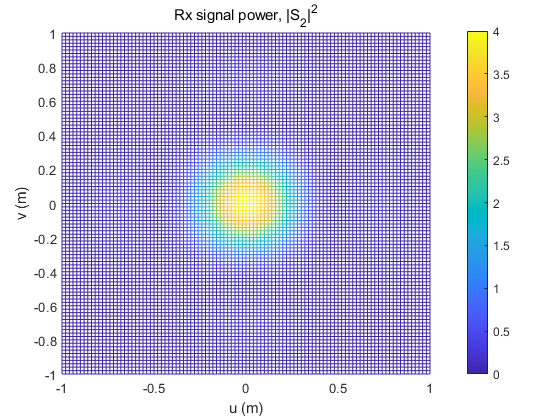}%
\label{fig.8_4}}
\caption{Signal transmission using two aligned metasurfaces. 
(a) Tx metasurface signal $S_1$ (b) Rx metasurface signal $S_2$  (c) Colormap of Tx metasurface signal $S_1$ (d) Colormap of Rx metasurface signal $S_2$}
\label{fig.8}
\end{figure*}

First, consider an environment with two aligned metasurfaces in Fig.~\ref{fig.2}.
On the Tx metasurface, a rectangular signal is transmitted as shown in Fig.~\ref{fig.8}(a) and (c),
where the center position of signal $(x_0,y_0)$ is $(0.2~m,0.2~m)$ and the signal width $(L_x,L_y)$ is $(0.2~m,0.2~m)$.
The total power $P(S_1)$ of the Tx metasurface signal is set to be 1.
In Fig.~\ref{fig.8}(b) and (d), the power of the Rx signal has the form of a sinc function as shown in (\ref{eq.55b}).
The Rx signal power $|S_2|^2$ becomes zero when $p=\pm 0.5~ m$ or $q=\pm 0.5~ m$. 
Note that $\frac{\lambda R}{L_x}=0.5~m$ and $\frac{\lambda R}{L_y}=0.5~m$.  
The total power $P(S_2)$ of the Rx metasurface is 0.92.
One reason for the difference between this power value and the result in (\ref{eq.9}) is that the area of the Rx metasurface is finite.

\begin{figure*}[!t]
\centering
\subfloat[]{\includegraphics[width=2.3in]{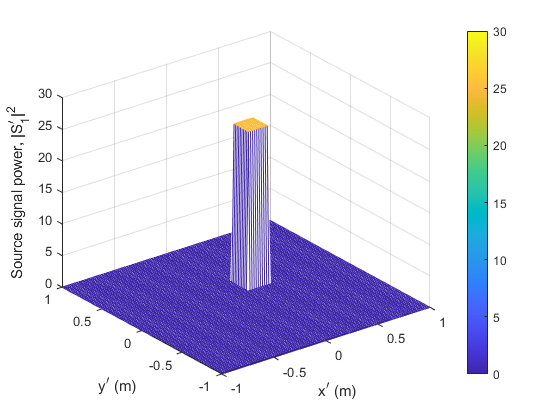}%
\label{fig.9_1}}
\hfil
\subfloat[]{\includegraphics[width=2.3in]{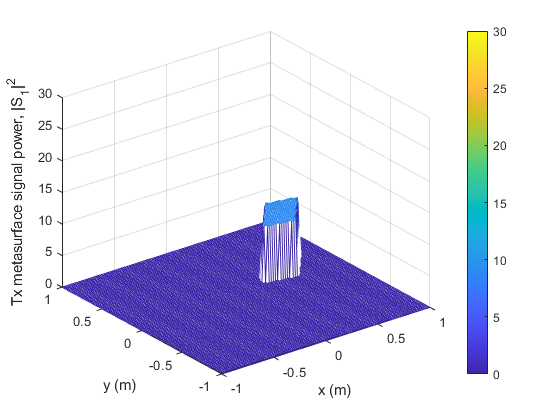}%
\label{fig.9_2}}
\hfil
\subfloat[]{\includegraphics[width=2.3in]{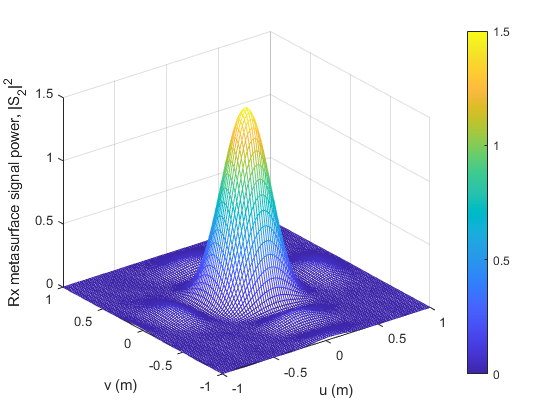}%
\label{fig.9_3}}
\hfil
\subfloat[]{\includegraphics[width=2.3in]{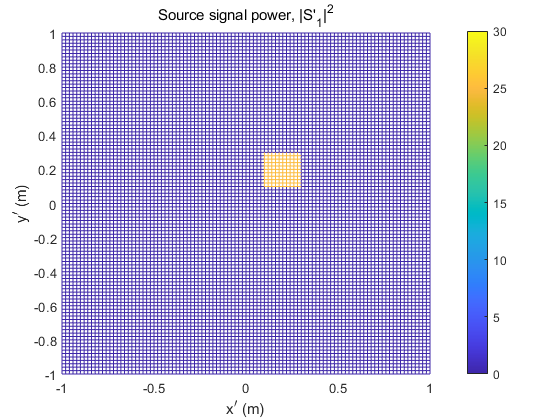}%
\label{fig.9_4}}
\hfil
\subfloat[]{\includegraphics[width=2.3in]{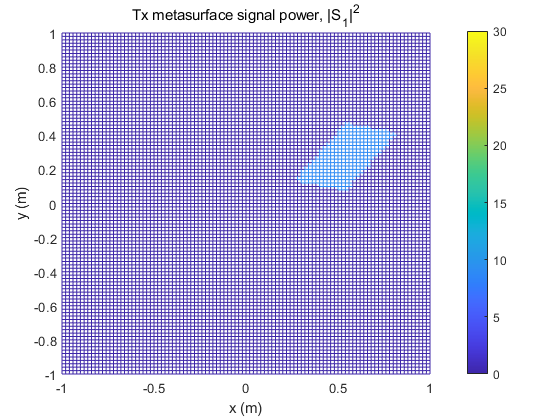}%
\label{fig.9_5}}
\hfil
\subfloat[]{\includegraphics[width=2.3in]{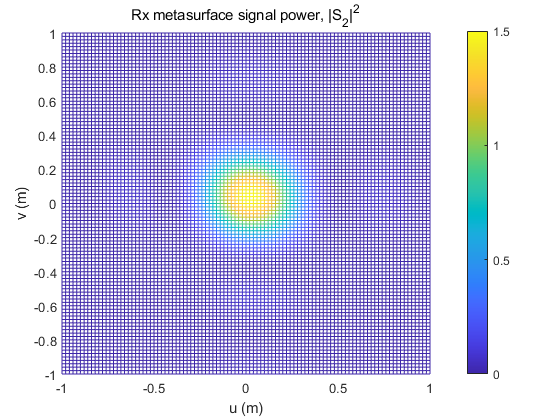}%
\label{fig.9_6}}
\caption{Signal transmission using two unaligned metasurfaces. 
(a) Source signal $S'_1$ (b) Tx metasurface signal $S_1$  (c) Rx metasurface signal $S_2$ 
(d) Colormap of source signal $S'_1$ (e) Colormap of Tx metasurface signal $S_1$ (f) Colormap of Rx metasurface signal $S_2$}
\label{fig.9}
\end{figure*}

Now, consider an environment with two unaligned metasurfaces in Fig.~\ref{fig.4}.
We set the unit direction vectors as follows:
\begin{subequations}
\begin{align}
&\hat{r} = \frac{1}{\sqrt{2}} \hat{y} -  \frac{1}{\sqrt{2}} \hat{z}, \label{eq.89a} \\
&\hat{u} = \frac{1}{2} \hat{x} + \frac{1}{\sqrt{2}} \hat{y} - \frac{1}{2} \hat{z}, 
 ~~\hat{v} = - \frac{1}{2} \hat{x} + \frac{1}{\sqrt{2}} \hat{y} + \frac{1}{2} \hat{z}, \nonumber \\ 
&\hat{w} = \frac{1}{\sqrt{2}} \hat{x} + \frac{1}{\sqrt{2}} \hat{z}. \label{eq.89b}
\end{align}
\end{subequations}
A rectangular source signal is transmitted as shown in Fig.~\ref{fig.9}(a) and (d),
where the center position of signal $(x_0,y_0)$ is $(0.2~m,0.2~m)$ and the signal width $(L_x,L_y)$ is $(0.2~m,0.2~m)$.
The total power $P(S'_1)$ is set to be 1.
On the Tx metasurface, the transformed signal is transmitted to compensate for the misalignment as shown in Fig.~\ref{fig.9}(b) and (e).
The total power $P(S_1)$ is 0.99, which is consistent with the result of (\ref{eq.35}).
Fig.~\ref{fig.9}(c) and (f) show similar patterns to Fig.~\ref{fig.8}(b) and (d)
since the signals $S'_1$ and $S_2$ have the Fourier transform relationship.
The total power $P(S_2)$ is $0.29$, compared to the theoretical value $0.35$ in (\ref{eq.38}).
The difference can be reduced if the size of the Rx metasurface is large enough.

\subsection{Spatial multimode transmission using three metasurfaces}
Now let's explore an environment using three metasurfaces.
The distance $R_1$ between the Tx metasurface and the RIS metasurface is $10~m$, 
and the distance $R_2$ between the RIS metasurface and the Rx metasurface is $5~m$.

\begin{figure*}[!t]
\centering
\subfloat[]{\includegraphics[width=2.3in]{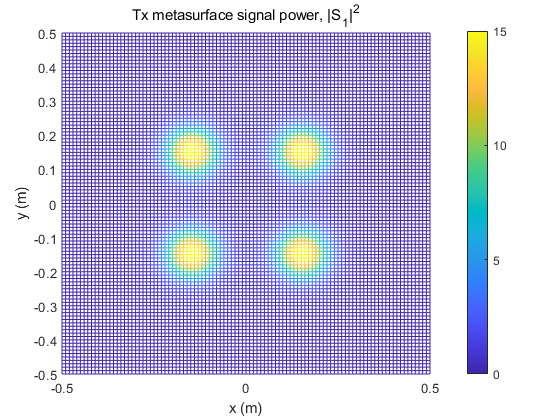}%
\label{fig.10_1}}
\hfil
\subfloat[]{\includegraphics[width=2.3in]{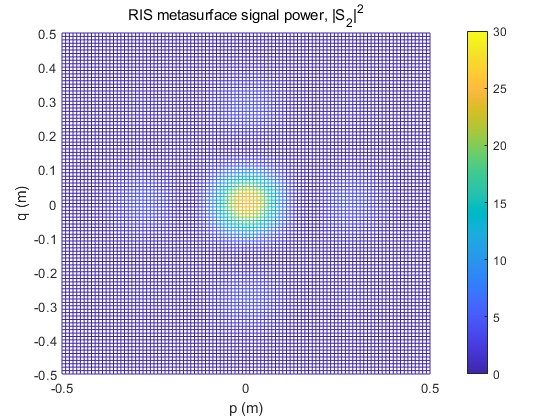}%
\label{fig.10_2}}
\hfil
\subfloat[]{\includegraphics[width=2.3in]{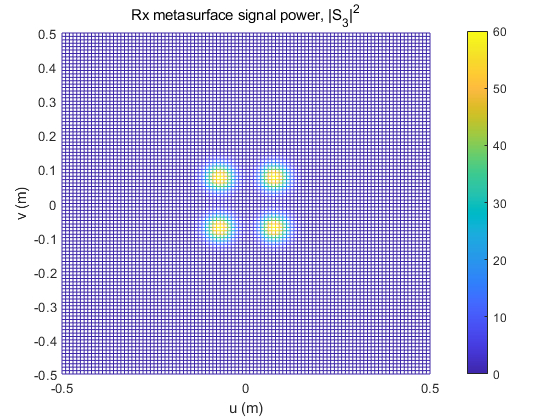}%
\label{fig.10_3}}
\caption{Signal transmission using three aligned metasurfaces. 
(a) Tx metasurface signal $S_1$ (b) RIS metasurface signal $S_2$ (c) Rx metasurface signal $S_3$}
\label{fig.10}
\end{figure*}

Consider a two-dimensional signal environment in Fig.~\ref{fig.5},
where three metasurfaces are aligned.
As shown in Fig.~\ref{fig.10}(a), four Gaussian signals in (\ref{eq.57}) are transmitted to the Tx metasurface,
where the center points of the Tx signal power are $(\pm 0.2~m, \pm 0.2~m)$ and $(\pm 0.2 m, \mp 0.2 m)$,
and the standard deviation $(\sigma_x, \sigma_y)$ of the Tx signal power is $(0.05~m, 0.05~m)$.
The total power $P(S_1)$ is set to 1.
The size of the RIS metasurface is set to $\left(\frac{\lambda R_1}{2 \sigma_x}, \frac{\lambda R_1}{2 \sigma_y} \right) = (1~m, 1~m)$. 
Four Gaussian signals overlap on the RIS metasurface as shown in Fig.~\ref{fig.10}(b).
The total power $P(S_2)$ is 1 as expressed in (\ref{eq.50}).
In Fig.~\ref{fig.10}(c), the center points of the Rx signal are $(\pm 0.1~m, \pm 0.1~m)$ and $(\pm 0.1~m, \mp 0.1~m)$, 
and the standard deviation of the power of the Rx signal is reduced to half compared to that of the power of the Tx signal, which is consistent with the results in (\ref{eq.59c}).
The total power $P(S_3)$ is 1 as shown in (\ref{eq.50}).

\begin{figure*}[!t]
\centering
\subfloat[]{\includegraphics[width=2.3in]{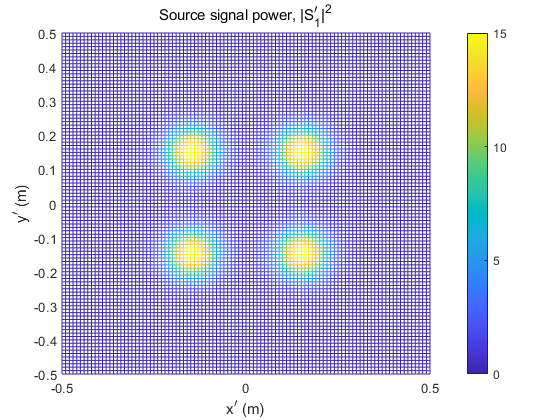}%
\label{fig.11_1}}
\hfil
\subfloat[]{\includegraphics[width=2.3in]{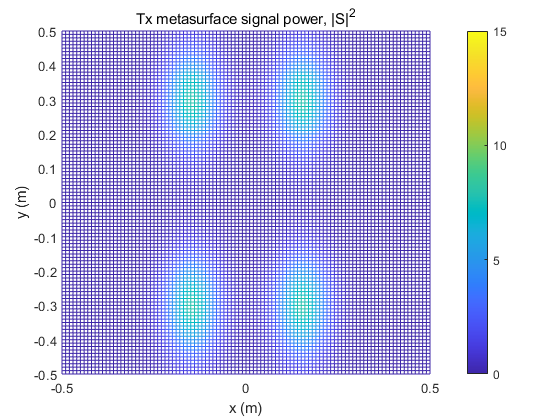}%
\label{fig.11_2}}
\hfil
\subfloat[]{\includegraphics[width=2.3in]{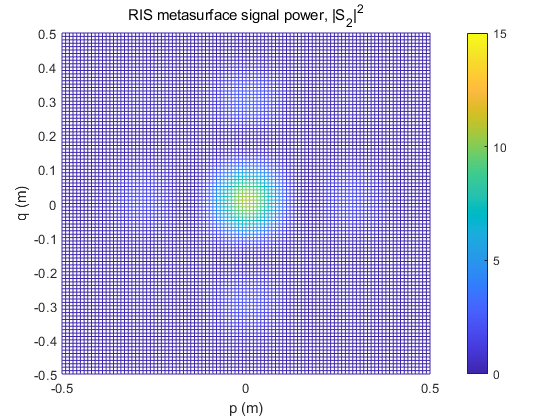}%
\label{fig.11_3}}
\hfil
\subfloat[]{\includegraphics[width=2.3in]{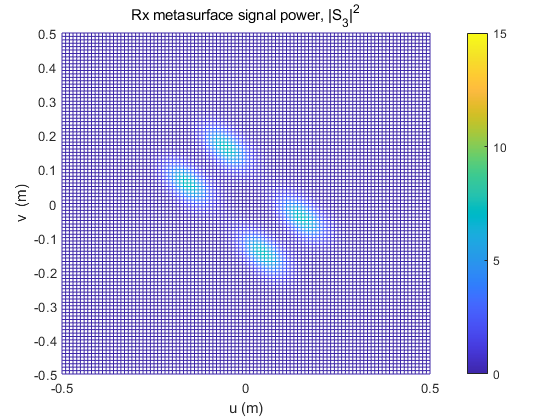}%
\label{fig.11_4}}
\hfil
\subfloat[]{\includegraphics[width=2.3in]{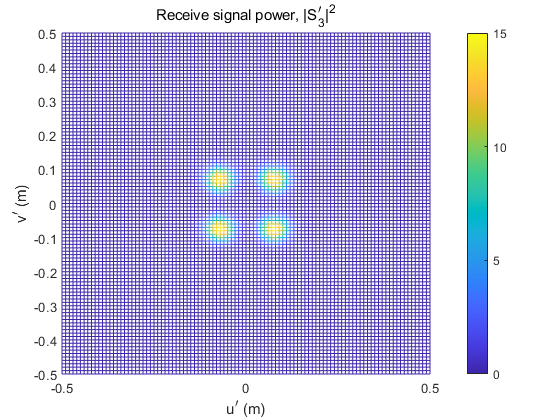}%
\label{fig.11_5}}
\caption{Signal transmission using three unaligned metasurfaces. 
(a) Source signal $S'_1$ (b) Tx metasurface signal $S_1$ (c) RIS metasurface signal $S_2$ (d) Rx metasurface signal $S_3$ (e) Received signal $S'_3$}
\label{fig.11}
\end{figure*}

Consider a two-dimensional signal environment in Fig.~\ref{fig.7},
where three metasurfaces are not aligned.
We set the unit direction vectors as follows:
\begin{subequations}
\begin{align}
&\hat{t} = \frac{1}{\sqrt{2}} \hat{x} + \frac{1}{\sqrt{2}} \hat{z}, ~~\hat{r} = - \frac{1}{\sqrt{2}} \hat{x} + \frac{1}{\sqrt{2}} \hat{z}, \label{eq.90a} \\
&\hat{p} = - \hat{z}, ~~\hat{q} = \hat{y}, ~~\hat{s} = \hat{x}, \label{eq.90b} \\
&\hat{u} = \frac{1}{\sqrt{2}} \hat{x} + \frac{1}{\sqrt{2}} \hat{y}, ~~\hat{v} = - \frac{1}{\sqrt{2}} \hat{x} + \frac{1}{\sqrt{2}} \hat{y}, ~~\hat{w} = \hat{z}. \label{eq.90c}
\end{align}
\end{subequations}
As shown in Fig.~\ref{fig.11}(a), four Gaussian signals in (\ref{eq.57}) are transmitted,
where the center points of the source signal power are $(\pm 0.2~m, \pm 0.2~m)$ and  
and $(\pm 0.2~m, \mp 0.2~m)$, and the standard deviation $(\sigma_x, \sigma_y)$ of source signal power is $(0.05~m, 0.05~m)$, respectively.
The total power of the source signal $P(S'_1)$ is set to 1.
The transformed signal is transmitted on the Tx metasurface to compensate for the misalignment, as shown in Fig.~\ref{fig.11}(b).
The total power $P(S_1)$ is 1 as expressed in (\ref{eq.85a}).
The area of the Tx metasurface signal $S_1$ is larger than the area of the source signal $S'_1$, 
but the total powers of the two signals are the same.
The size of the RIS metasurface is set to $\left(\frac{\lambda R_1}{2\sigma_x}, \frac{\lambda R_1}{2\sigma_y} \right) = (1~m, 1~m)$. 
As shown in Fig.~\ref{fig.11}(c), four Gaussian signals overlap on the RIS metasurface.
The total power $P(S_2)$ is 0.5 as shown in (\ref{eq.86a}).
On the Rx metasurface in Fig.~\ref{fig.11}(d), the receive signal is still distorted due to the misalignment.
The total power $P(S_3)$ is $0.25$ as shown in (\ref{eq.86b}).
In Fig.~\ref{fig.11}(e), the center points of the Rx signal are $(\pm 0.1~m, \pm 0.1~m)$ and $(\pm 0.1~m, \mp 0.1~m)$, 
and the standard deviation of the Rx signal power is reduced to half compared to that of the Tx signal power.
The total power $P(S'_3)$ is $0.25$ as shown in (\ref{eq.85b}).

Fig.~\ref{fig.11}(c) shows a pattern similar to Fig.~\ref{fig.10}(b)
since the signals $S'_1$ and $S_2$ have a Fourier transform relationship. 
Also, Fig.~\ref{fig.11}(e) shows the similar pattern to Fig.~\ref{fig.10}(c)
since the signals $S_2$ and $S'_3$ have the Fourier transform relationship.

\section{Conclusions} \label{section7}

In this paper, we proposed a spatial multimode transmission technology for efficient use of spatial resources for 6th generation mobile communications.
By using a metasurface that is sufficiently larger than the wavelength, it is possible to obtain a Fourier transform relationship between the transmitted and received metasurface signals through phase control on the metasurface.
Especially, we proposed a method to obtain the Fourier transform relationship through conversion of the transmitted and received signals
when the Tx metasurface and the Rx metasurface are not aligned.
We also showed that by using three metasurfaces, multiple streams can be transmitted and received with low complexity.
Analysis of the total power of the metasurface and the maximum number of transmittable spatial multimodes were also performed.

So far, frequency resources have been mainly used in the time domain in mobile communications, but it is expected that systematic use of the spatial domain will be possible by utilizing metasurfaces at high frequencies.
Since the 4th generation mobile communication technology, 
many attempts have been made to obtain the benefits of multiple streams in the spatial domain through MIMO, 
but it has not been easy to obtain sufficient ranks. 
In the 6th generation mobile communication technology, 
we can achieve predictable capacity gains of the space resources instead of an unpredictable scattering gain for MIMO transmission.
Additionally, the communication technology to optimally utilize frequency resources in the time domain can be applied to spatial multimode transmission technology 
to optimally utilize metasurface resources in the spatial domain.


\section*{Acknowledgment}

This work was supported by Institute of Information \& communications Technology Planning \& Evaluation (IITP) grant funded by the Korea government(MSIT) (2019-0-00500, RS-2019-II190500, Development of spatial multi-mode transmission technologies for 6G mobile communications )

\appendix[Proof of (\ref{eq.37})]
\setcounter{equation}{0}
\numberwithin{equation}{section}
\begin{proof}
Let $\bar{u}'$ and $\bar{v}'$ be the corresponding vectors 
when the two orthogonal unit vectors, $\hat{u}$ and $\hat{v}$ are projected onto a plane perpendicular to $\hat{r}$. 
Also, let $\bar{u}''$ and $\bar{v}''$ be the corresponding vectors 
when the two vectors $\bar{u}'$ and $\bar{v}'$ are projected onto a plane perpendicular to $\hat{z}$. 
Also, let $D$ be the area of the parallelogram composed of two vectors of $\bar{u}''$ and $\bar{v}''$.

Note that the following hold:
\begin{subequations}
\begin{align}
&\bar{u}' = \hat{u} - (\hat{r} \cdot \hat{u}) \hat{r} = \hat{u} - a_{ru} \hat{r},  \label{eq.A1a} \\
&\bar{v}' = \hat{v} - (\hat{r} \cdot \hat{v} ) \hat{r} = \hat{v} - a_{rv} \hat{r}.  \label{eq.A1b}
\end{align}
\end{subequations}
Also, $\bar{u}''$ can be expressed as:
\begin{align}
\bar{u}'' 
& = \bar{u}' - (\hat{z} \cdot \bar{u}') \hat{z} \nonumber \\
& = (\hat{u} - a_{ru} \hat{r}) - (a_{zu} - a_{rz} a_{ru}) \hat{z}  \nonumber \\
& = ( a_{xu} \hat{x} + a_{yu} \hat{y} + a_{zu} \hat{z} ) - a_{ru} ( a_{rx} \hat{x} + a_{ry} \hat{y} + a_{rz} \hat{z} ) \nonumber \\
& ~~ - ( a_{zu} - a_{ru} a_{rz} ) \hat{z} \nonumber \\
& = ( a_{xu} - a_{rx} a_{ru} ) \hat{x} + ( a_{yu} - a_{ry} a_{ru} ) \hat{y} \nonumber \\
& = b_{xu} \hat{x} + b_{yu} \hat{y}. \label{eq.A2}
\end{align}
Using the above approach, we can obtain the following:
\begin{equation}
\bar{v}'' = b_{xv} \hat{x} + b_{yv} \hat{y}. \label{eq.A3}
\end{equation}
Therefore, the area $D$ of the parallelogram created by the vectors $\bar{u}''$ and $\bar{v}''$ is as follows:
\begin{equation}
D = \left\| \begin{matrix} b_{xu} & b_{xv} \\ b_{yu} & b_{yv} \\ \end{matrix} \right\| = || T ||.    \label{eq.A4}
\end{equation}

On the other hand, from (\ref{eq.A1a}) and (\ref{eq.A1b}),
\begin{subequations}
\begin{align}
| \bar{u}' |^2 = 1 - ( a_{ru} )^2 , \label{eq.A5a} \\
| \bar{v}' |^2 = 1 - ( a_{rv} )^2 . \label{eq.A5b}
\end{align}
\end{subequations}
Let $\phi$ be the angle formed by $\bar{u}'$ and $\bar{v}'$. 
Then, the area of the plane formed by $\bar{u}'$ and $\bar{v}'$ is as follows:
\begin{align}
|\bar{u}'| |\bar{v}'| |\sin(\phi) | 
&= \sqrt{ |\bar{u}'|^2 |\bar{v}'|^2 - ( \bar{u}' \cdot \bar{v}')^2 }  \nonumber \\
& = \sqrt{ [ 1- (a_{ru})^2] [ 1- (a_{rv})^2] - (a_{ru} a_{rv})^2} \nonumber \\
& = | a_{rw} |,   \label{eq.A6}
\end{align}
since $a_{ru}^2 + a_{rv}^2 + a_{rw}^2 = 1$.
We can also show that 
when the unit area in the plane orthogonal to the vector $\hat{r}$ is projected into the plane with unit vectors $\hat{x}$ and $\hat{y}$,
the projected area becomes $|a_{rz}|$.
Therefore, the following is derived:
\begin{equation}
D = | a_{rz} a_{rw} |. \label{eq.A7}   
\end{equation}
\end{proof}

\end{document}